\documentclass[prc,twocolumn,secnumroman,tightenlines,amssymb,aps,nobibnotes,superscriptaddress,balancelastpage,floatfix,nofootinbib]{revtex4}
\usepackage{soul}  

\usepackage{amsmath}
\usepackage{soul}
\usepackage{float}
\usepackage{lipsum}
\usepackage{graphicx}
\usepackage{multirow}           
\usepackage{color}
\usepackage{array}
\usepackage{booktabs}

\DeclareRobustCommand\substyle{\name@idx{document substyle}}
\DeclareRobustCommand\classoption{\name@idx{document class option}}
\DeclareRobustCommand\classname{\name@idx{document class}}
\def\name@idx#1#2{{\ttfamily#2}
\index{#2\space#1=\string\ttt{#2}\space#1}\index{#1>#2=\string\ttt{#2}}}

\DeclareTextFontCommand{\rb}{\color{red}\bfseries}

\begin{document}







\title{ Impact of new results from the ultraperipheral collision on modeling the proton and neutron emission in photon-induced nuclear processes}

\author{P. Jucha}
\email{Pawel.Jucha@ifj.edu.pl}
\affiliation{Institute of Nuclear Physics PAN, ul.\,Radzikowskiego 152, Pl-31342 Krak\'ow, Poland}

\author{K. Mazurek}
\email{Katarzyna.Mazurek@ifj.edu.pl}
\affiliation{Institute of Nuclear Physics PAN, ul.\,Radzikowskiego 152, Pl-31342 Krak\'ow, Poland}

\author{A. Szczurek}
\email{Antoni.Szczurek@ifj.edu.pl}
\affiliation{Institute of Nuclear Physics PAN, ul.\,Radzikowskiego 152, Pl-31342 Krak\'ow, Poland}
\affiliation{Institute of Physics, Faculty of Exact and Technical Sciences, University of Rzeszów, Pigonia 1 St., 35-310 Rzeszów, Poland}

\author{K. Pysz}
\email{Krzysztof.Pysz@ifj.edu.pl}
\affiliation{Institute of Nuclear Physics PAN, ul.\,Radzikowskiego 152, Pl-31342 Krak\'ow, Poland}


%
\date{\today}

\begin{abstract}
The ultrarelativistic collisions of heavy ions provide rich 
spectrum of possibilities to discuss the response of the nucleus 
to photons. Newly published
neutron and proton multiplicities measured in the ALICE experiment in
ultraperipheral collisions allow investigating the influence of the
electromagnetic fields on colliding nuclei for the
$^{208}$Pb+$^{208}$Pb at $\sqrt{s_{NN}}$=5.02~TeV. The theoretical
predictions are done within hybrid model including equivalent photon
approximations (EPA), GiBUU modeling of pre-equilibrium 
processes and generation of the excited nuclear remnants, which decay is modeled 
by statistical approach: GEM2 or GEMINI++. The cross-sections of the mass-charge distributions of nuclear remnants as well as the neutron, proton and other charged particle multiplicities are estimated. We concentrate on production of protons and isotopes coming from the electromagnetic dissociation. Special attention is devoted to emission of a single proton.
 It is explained in the paper that cross section for $1p$ emission is very close to maximal available one based on reactions of photon with individual nucleons (protons or neutrons) inside the nucleus.

\end{abstract}


%


\maketitle


\section{Introduction}
\label{Sect.I}

The particle multiplicities are one of the basic observables measured 
in many low and high energy collision experiments. In ultrarelativistic
heavy-ion reactions this observable is not straightforward
as the emitted particles go in very forward or very backward directions.
 The multiplicity of neutrons in ultraperipheral collisions (UPC) of $^{208}$Pb + $^{208}$Pb published by the ALICE Collaboration \cite{ALICE502a} was successfully described in \cite{Jucha2025, Jucha2025a}.
For this aim the photon flux 
was calculated using
the equivalent photon approximation (EPA) \cite{b-space-EPA}.
To calculate the photon interaction with the nuclear system, the phenomenological two-component model (TCM) \cite{Jucha2025} was proposed.
In this approach the excitation of a nucleus induced by absorption of (single or multiple)
photons has been assumed to be a combination of two exponential functions
with one free parameter. The remnant of the photon-nucleus interaction with calculated in this way excitation
energy has been further processed with statistical models EMPIRE \cite{empire}, GEMINI++ \cite{gemini}, HIPSE \cite{hipse1},
and their combinations. Theoretical neutron multiplicities
were obtained from the combination of these models.

The best agreement with the ALICE data was observed
for simulations with a simple model proposed by us called TCM + GEMINI++.
The details of the photon-nucleus collision process and the energy released during the intranuclear cascade
as a kinetic energy in pre-equilibrium emission of neutrons and other 
particles were disregarded in the model.

Recently, the ALICE collaboration presented also results for proton
emission \cite{ALICE_protons} based on the use of so-called proton
zero-degree calorimeters (ZDC). As will be discussed in our paper
the new results shed new light on nuclear photoproduction processes.

The present work extends previous considerations
by including a microscopic description of the photon-nucleus collision process for photon energy above 200~MeV.
This approach gives insight into yields of pre-equilibrium emission
of nucleons and pions, and thus an explicit distribution of excitation energies
of UPC remnants. The fate of the latter can be examined with the deexcitation models for thermalized nuclei.

While in the previous paper \cite{Jucha2025,Jucha2025a} the contribution of low-energy component
of photon flux (below 200 MeV) was examined, here, the emphasis is put
on the contribution of photons with energies extending from 200 MeV to almost 1 TeV. 
In summary, the hybrid model discussed here consists
of three main parts:\\
(1) Calculation of the photon flux surrounding the colliding Pb nucleus (EPA model).\\
(2) Estimation of photon-Pb interaction in the pre-equilibrium phase by
excitation energy of remnants, and mass-charge distribution of the hot
spectator (TCM or HIPSE for energy range 20-200~MeV, and GiBUU 
for more energetic $\gamma$-rays, $E_{\gamma} >$ 200 MeV).\\
(3) De-excitation of the collision remnants created in a pre-equilibrium phase, with the help of models: GEM2 or GEMINI++ to obtain the particle multiplicities and mass-charge distribution of the final products.

The paper is organized as follows: the details of the theoretical approaches used are shown in Sect.~\ref{Sect.II}; in Sec.~\ref{Sect.III} results for mass-charge distribution of Pb remnants as well as the neutron and proton multiplicity are presented. The Section~\ref{Sect.IV} contains a comparison of the obtained results
with the data measured by the ALICE experiment and a detailed discussion of the processes participating in the total cross section in photon-nucleus collisions.
The conclusions based on the data close the paper.

\section{Theoretical models}
\label{Sect.II}
\subsection{Equivalent photon approximation model and photon flux}
The electromagnetic interaction caused by moving charged nuclei generates the photon flux, which is
calculated using the equivalent photon approximation (EPA) \cite{b-space-EPA}. 
The analytic formula for a point-like nucleus could be written as e.g. in~\cite{Jackson}:

\begin{equation}
N(\omega, b) = \frac{\alpha Z^2}{\pi^2} \frac{u^2}{\beta^2 \omega b^2} \left( K^2_1(u) + \frac{1}{\gamma^2} K^2_0(u) \right).  
\label{Eq.1}
\end{equation}
For extended charge, one has:

\begin{equation}
            N(\omega,b) = \frac{Z^2 \alpha_{em}}{\pi^2 \beta^2}\frac{1}{\omega b^2} \times \left|
            \int d\chi \hspace{3pt} \chi^2 
            \frac{F(\frac{\chi^2+u^2}{b^2})}{\chi^2+u^2}J_1(\chi) \right|^2 .
\label{Eq.2}
\end{equation}

\noindent Above $\omega$ is photon energy and $b$ is transverse distance of photons from the emitting nucleus. The photon energy in the rest frame of the absorbing nucleus gives $\omega=E_{\gamma}$. In the case of our reaction we need a characteristics of the fluxes in the rest frame of the target nucleus. At the LHC energies $\beta\approx 1$. Since $\gamma$ is very large, the second term in Eq.~(\ref{Eq.1}) practically vanishes. The electromagnetic form factor of the nucleus $F(\frac{\chi^2+u^2}{b^2})$ depends on $u = \frac{\omega b}{\gamma\beta}$, $\gamma=\frac{1+\beta^2}{1-\beta^2}$, $\chi$ is an auxiliary dimensionless variable related to the photon transverse momentum via the relation $\chi = k_\perp b$. In the realistic case $F(\frac{\chi^2+u^2}{b^2})$ is calculated as a Fourier transform of the nucleus charge density.

The EPA method allows us to estimate the impact parameter space distribution of photons produced by fast moving charged sources. In Eq.~(\ref{Eq.2}), $N(\omega,b)$ means rather $\frac{dN(\omega,b)}{d\omega d^2b}$  to shorten other more complicated formulas. Photons are typically produced with energies below 20~MeV, but there is a sizable probability of generating highly energetic $\gamma$-rays.
This energetic quasi-real photons appear for the intra-nuclear impact 
parameter $b = R_{A_1} + R_{A_2}>$ 14~fm, where UPC take place.
The partner nucleus absorbs such photons and gets excited. The excitation energy range is typical for nuclear processes such as $\gamma$-ray and particle emission, fission, or even multifragmentation. Excitation of low-mass nucleon resonances can be involved. \footnote{$\Delta^+ or \Delta^0$ are the most representative examples.}
\subsection {\label{subsec:gibuu} Models of collision process}
Microscopic models of nuclear collisions, which include photons as a projectile,
are scarce. Among them the RELDIS code has to be mentioned \cite{PhysRevC.60.044901}, but unfortunately
it is not accessible to us. We decided to employ the broadly used GiBUU model \cite{BUSS20121}.
It is a model of hadron, nucleus and lepton collisions with
various targets. Large scope of hadron physics as well as a Heavy Ion (HI) collision phenomena
are implemented at very broad energy range. The good predictive power of GiBUU model for various fields
of its implementation was verified many times
(see, e.g. \cite{ALIAGASOPLIN2025109553,HADES:2023ffa,HADES:2023sre,PhysRevC.85.024614}).

The GiBUU model is unique because it also includes the photon-N and photon-A interaction.
It is done explicitly for a real photon at low energies (below $\approx$ 2 GeV) and via virtual
photon created in lepton interaction at higher energies.

The GiBUU model utilizes the so-called Boltzmann-Uehling-Uhlenbeck transport equation.
The solution of BUU equation is performed by Monte Carlo simulations of motion of involved particles.
It is done with sets of test particles, where {\it n} of test particles replace each
of a reaction constituent. For the resulting distributions, the contributions from all test particles
are summed up with appropriate weights.

In case of reactions where nuclei are involved,
the BUU model calculations are able to produce the distributions of mesons
and single nucleons, both emitted and those forming a remnant.
The remnant nucleus of the intranuclear cascade can be defined by
its mass, charge, energy, and excitation energy.

Due to a lack of a mechanism that allows for particle correlations,
the simulation with the GiBUU must be restricted only to emission
of single nuclear products.

For the simulations with the use of GiBUU, in general, the standard parameters proposed 
by the authors of the model, were used \cite{BUSS20121}. However, since in the explored energy ranges two different approaches
for the photon-induced reaction are demanded (real photon at low energy and virtual photon at high energy),
the simulation parameters have been a bit adjusted in order to obtain a smooth transition between
the low- and high-energy results. The estimated uncertainty of
simulation results due to a variation in model parameters
is estimated for 20\%.

In our recent article \cite{Jucha2025} we showed that the excitation
energy of nucleus in the equilibrium state is not equal to the absorbed
photon energy. To describe this effect, we proposed a phenomenological
approach called Two-Component Model (TCM). In this simple model the 
nucleus excitation energy is encoded in the function

\begin{equation}
            P(E_{exc}; E_{\gamma}) = 
                 c_1(E_{\gamma}) \delta \left( E_{exc}-E_{\gamma} \right)
               + c_2(E_{\gamma}) / E_{\gamma} \; ,
               \label{two-component_model}
\end{equation}

where
$P(E_{exc}; E_{\gamma})$ (see \cite{Jucha2025}) is a conditional distribution in $E_{exc}$ populated at a given photon energy $E_{\gamma}$ and
\begin{eqnarray}
c_1(E_{\gamma}) &=& \mathrm{exp}\left( -E_{\gamma}/E_0 \right) \; ,
\label{eq:c1_E}
\\
c_2(E_{\gamma}) &=& 1 - \mathrm{exp} \left( -E_{\gamma}/E_0 \right) \; .
\label{eq:c2_E}
\end{eqnarray}

\noindent The $E_0$ parameter was set to the value of 50 MeV. Using the
probabilistic formula for GEMINI++ results (see Sec.~\ref{subsec:gem}), one is able to describe the
experimental data from photon-nucleus interactions with evaporation of
a given number of neutrons \cite{Jucha2025}.

The EMPIRE framework is a modular system of nuclear reaction codes. It
combines various nuclear models to permit description of nuclear
collisions in a broad energy range (keV- hundreds of MeV) and different incident
particles (neutron, 
proton, light and even heavy nuclei or photon).
The code accounts for the major nuclear reaction mechanisms, including direct (ECIS03, CCFUS models), pre-equilibrium (ORION + TRISTAN model). The compound nucleus is treated by
NVWY multi-step compound approach or by either a pre-equilibrium exciton model with cluster emission
(PCROSS) or by another one with full angular momentum coupling (DEGAS). The compound
nucleus decay is described by the full-featured Hauser-Feshbach model with $\gamma$-cascade and width 
fluctuations. The fission channel takes into account transmission through a
multiple-humped fission barrier with absorption in the wells. The fission probability is derived in
the WKB approximation using the optical model. 
The several data bases with experimental data or pre-calculated results
 with a microscopic model based on Hartree-Fock-Bogolyubov (HFB) single-particle level schemes with collective enhancement, nuclear masses, optical model parameters, ground state deformations, discrete levels and decay schemes, level densities, fission barriers, moments of inertia and $\gamma$-ray strength functions are included.

Another well-known collisions model relevant for particle+nucleus 
collisions is Heavy Ion Phase-Space Exploration (HIPSE) \cite{hipse1}
devoted to intermediate energy range.

The HIPSE model describes the heavy-ion collision in several stages: from the contact point, by reaggregation of the constituents of overlapping nuclei, to de-excitation of prefragments. This last stage is performed within the state-of-the-art statistical approach GEMINI++ \cite{gemini}. 
In general, it was used with success to describe reaction with collision
energies  15~--~100~MeV/nucleon. The reaction n+$^{207}$Pb is chosen in
\cite{Jucha2025}, as the neutron interactions seem to be the only possible approximation of the photon interaction. Strictly speaking the HIPSE 
does not allow for a photon-induced reaction. 

\subsection {\label{subsec:gem} De-excitation models}
There exist several models, both statistical and dynamical, which describe the de-excitation of a hot rotating nucleus.

The Weisskopf-Ewing formalism \cite{Weisskopf-Ewing} is used 
for particle evaporation from the excited nucleus
in energy equilibrium in the General Evaporation Model (GEM2) \cite{Furihata1,Furihata2}.
The density of states is given by the Gilbert-Cameron formula \cite{Gilbert_Cameron}
with modification as in the LAHET code \cite{LAHET}. The relevant reaction cross sections are parametrized.
The kind of particles emitted is determined by sampling its probability distribution.
Coulomb barrier is taken into account. The emission of both the stable as well as the excited ejectiles is considered.
The angular distribution of emitted ejectiles is isotropic in the center-of-mass system of the parent nucleus.
The angular momenta of the involved particles are not taken into account.

 One of the most popular is GEMINI++ statistical code~\cite{gemini,Cie15}, where the evaporation process is described by the Hauser-Feshbach formalism~\cite{hauser}. The spin,
total angular momentum, and orbital angular momentum of the evaporated 
particle are taken into account in each step. The total energy is 
conserved. The rotation is treated assuming the rigid body model. 
The level density parameter formula includes the shell correction to 
the liquid-drop mass and depends on the excitation energy. 
The separation energies $B_{i}$, nuclear masses, shell and pairing 
corrections are taken from Ref.~\cite{moller}.  

In the current work the Monte Carlo \textsc{GEMINI++} calculations assume that the excited nucleus is formed with zero angular momentum, which is a good approximation for the photon-induced reaction. At high energies protons or neutrons may escape in a preequilibrium process which leads to $E_{exc} < E_{\gamma}$. This strongly limits the excitation energy of the residual nucleus. Only for energy $E_{\gamma} <$ 100 MeV photon is totally absorbed and may lead to total angular momentum of the excited  nucleus. It is easy to estimate the total angular momentum of the nucleus as $L_{max} \approx 100~MeV~R_A$ where $R_A$ is the  absorbing nucleus radius. So the maximal angular momentum is about 10~$\hbar$.
We have made an additional calculation for L=10~$\hbar$ and obtained almost no effect on proton and neutron multiplicity.
Even if the initial angular momentum is zero, the angular momentum of nuclei in the evaporation cascade is not zero.
\begin{table}[!bt]
\caption{The selected features of the models used in the current calculations.}
	\begin{tabular}{l | c c c c }
		&&&\\Approach & reaction  & pre/post &  $E_\gamma$ [MeV]  & physics  \\ &&&\\\hline &&&\\
		TCM       & $\gamma$ + nucl.				   & pre-post 	& 8 - 200		& phenomenology	 \\
		HIPSE     & n + nucl.  & pre  			& 20 - 500		& nuclear \\
		GiBUU     & $\gamma$ + nucl. & pre  			& 200 - 10$^6$	& part. \& hadr. \\
		GEMINI++  & nucl.(E$^{*}$) 	   & post 			& 8 - 200		& nuclear \\
		GEM2      & nucl.(E$^{*}$) 	   & post 			& 8 - 2000		& nuclear \\ 
		EMPIRE    & $\gamma$ + nucl. & pre-post	& 8 - 900		& nucl. \& hadr. \\
		&&&\\\hline
	\end{tabular}
    \label{tab:models}
\end{table}

For a short summary, Table~\ref{tab:models} gives the selected features for the approaches used in the current calculation, to underline the differences between them. The GEMINI++ and GEM2 are the deexcitation codes thus as an input need only hot nuclei with a given excitation energy. The main distinction between GiBUU/EMPIRE and HIPSE is the possibility of the description of the $\gamma$+nucleus reactions and accounting also hadronic and partonic physics.

\section{Results}
\label{Sect.III}

The GiBUU approach provides cross sections for various reactions. The
total cross section is approximately similar to the absorption cross
section \cite{Mainz}. Fig.~\ref{Photo_abs} shows that for photon
energies above 200~MeV, the theoretical (estimated by GiBUU) and the experimental
photoabsorption cross sections agree.

\begin{figure}[!bt]
\includegraphics[width=8cm]{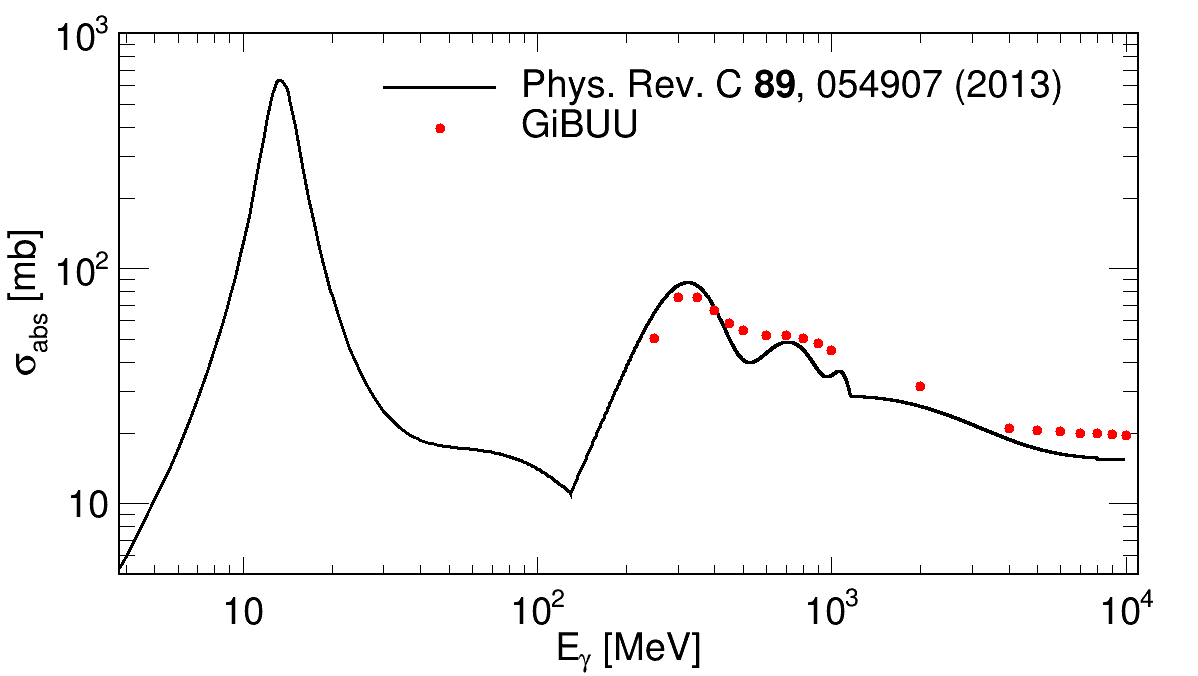}
\caption{Photoabsorption cross section for $\gamma$+Pb reaction in energy range 10 MeV- 100 GeV (full line) compared with the total cross section obtained with the GiBUU model (dots). 
}
\label{Photo_abs}
\end{figure}
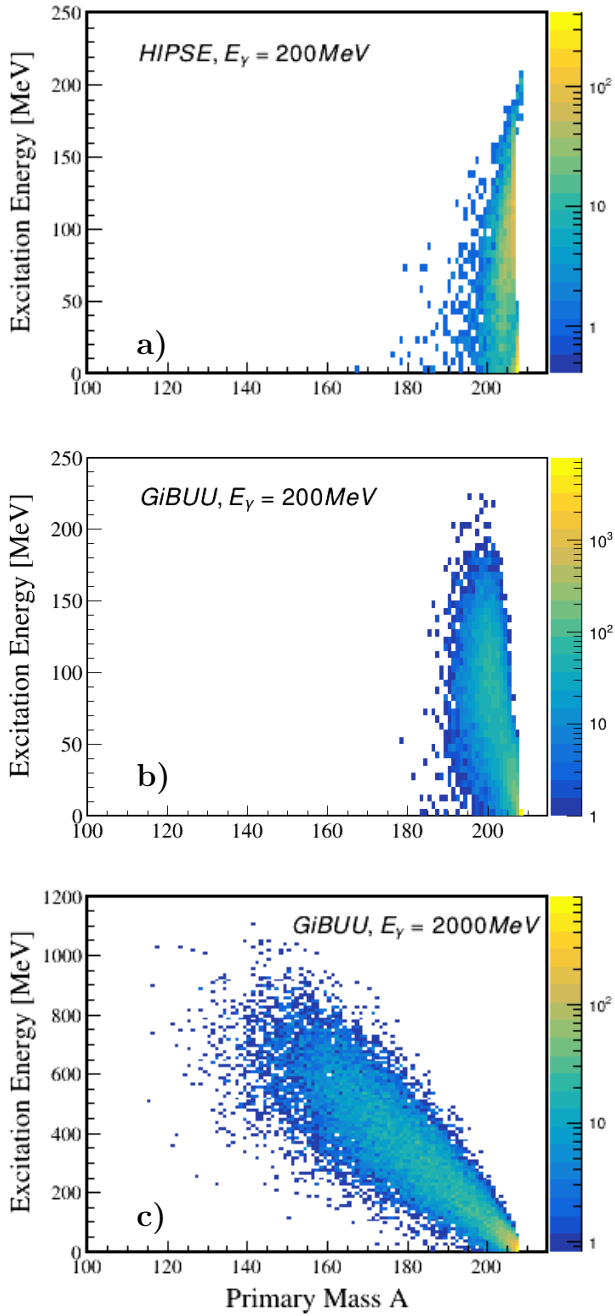
\begin{figure}[!bt]
\setlength{\unitlength}{0.1\textwidth}
\begin{picture}(4,10.)
\put(-0.5,6.50){\includegraphics[width=3.50in]{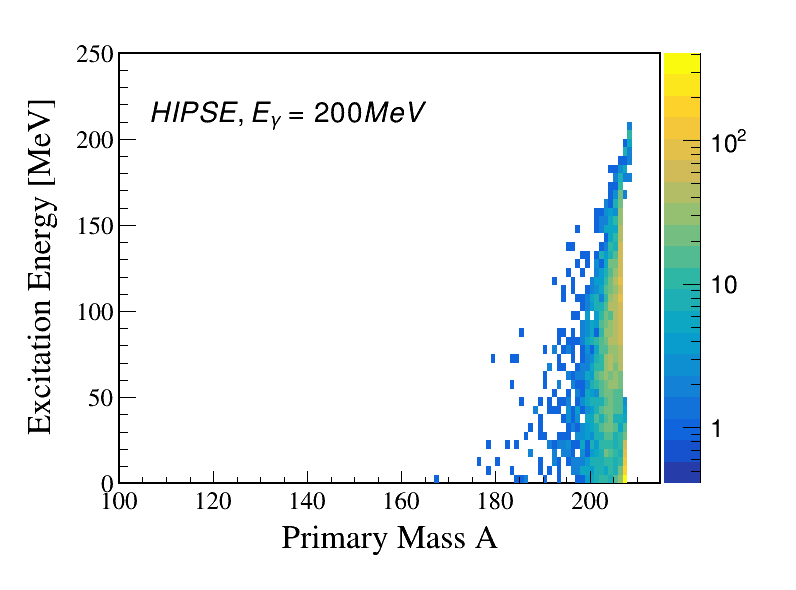}}
\put(0.6,7.3){\bf \large a)} 
\put(-0.5,3.22){\includegraphics[width=3.50in]{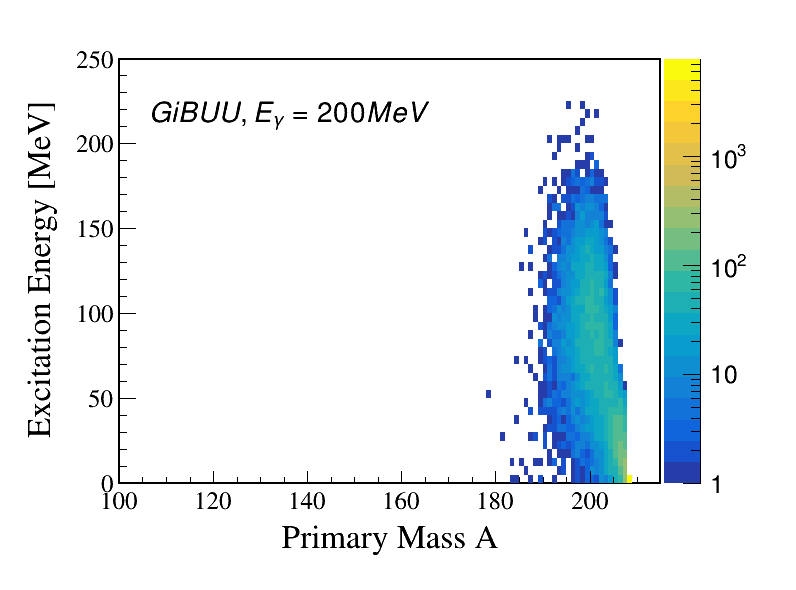}}
\put(0.6,4.10){\bf \large b)}
\put(-0.5,0){\includegraphics[width=3.50in]{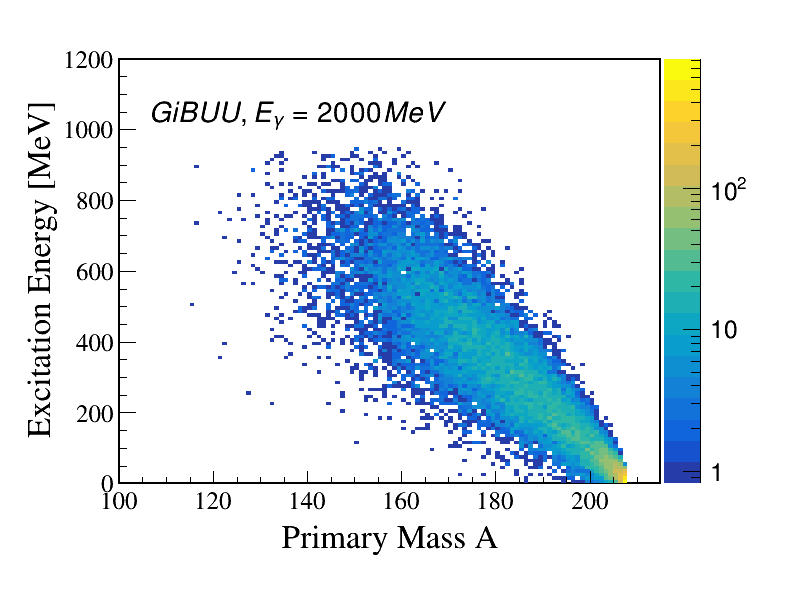}}
\put(0.6,0.85){\bf \large c)} 
 \end{picture}
\caption{Distribution of the primary mass and its excitation energy
  predicted in the naively adopted HIPSE ($E_{\gamma}$=200~MeV - top), 
and GiBUU ($E_{\gamma}$=200~MeV - middle and $E_{\gamma}$=2000~MeV - 
bottom.)
}
\label{excitation}
\end{figure}

The collision of the photon and nucleus is very interesting process and
many approaches try to describe it. The mechanisms of photon-nucleus interaction are energy-dependent. In the UPC, different photon energies and therefore various mechanisms come into play. One has to consider the situation when photons only slightly excite nucleus (giant dipole resonances), photon knock-out of one nucleon, excitation of nucleon resonances (mainly $\Delta$), many exclusive processes on individual nucleons and finally partonic reactions on nucleons (nucleon break-up).
Afterwards, in all cases, various particles can be emitted during de-excitation, by means of evaporation or by more violent processes.     

In most cases of interest of this work, the result of a collision considered has to be a highly excited nuclear system. There are various approaches to estimate its excitation energy. One of them is the energy
balance, where part of the photon energy is taken as kinetic energy of
evaporated particles and the Q-values. This procedure is applied, for
example, in the HIPSE model
(Fig.~\ref{excitation} a). The HIPSE model is usually applied for reactions up to the Fermi energy, thus during the reagregation stage the energy and momentum conservation laws are fulfilled, allowing to build clusters. In the fragmentation energy regime the internal energy forbids creation of bigger systems. During the thermalization phase, the full internal energy is divided among the formed prefragments, allowing an estimation of their excitation energy.

In the case of approaches used for description of nucleus behavior when the
nuclear processes are saturated, it is very difficult to discuss about
prefragments. For our purposes, the result of the collision $\gamma$ +
Pb was translated into prefragments assuming that nucleons close to each
other in space could form clusters and even heavier nuclei. The energy
conservation law is not really the best method to estimate the
excitation energy of such a system. Thus, assuming that one missing
particles takes away on average 13~MeV the excitation energy left in the
compound system is extracted from photon energies. In the GiBUU approach, the sum of energies taken by
missing nucleons (vacancies) generate the excitation energy
(Fig.~\ref{excitation} b, c). 
Very high excitation energies could be generated within the GiBUU model.
The HIPSE and GiBUU give different excitation energy distributions as it is
shown in Fig.~\ref{excitation} a, b. 

\begin{figure}[!bt]
\includegraphics[width=9.5cm]{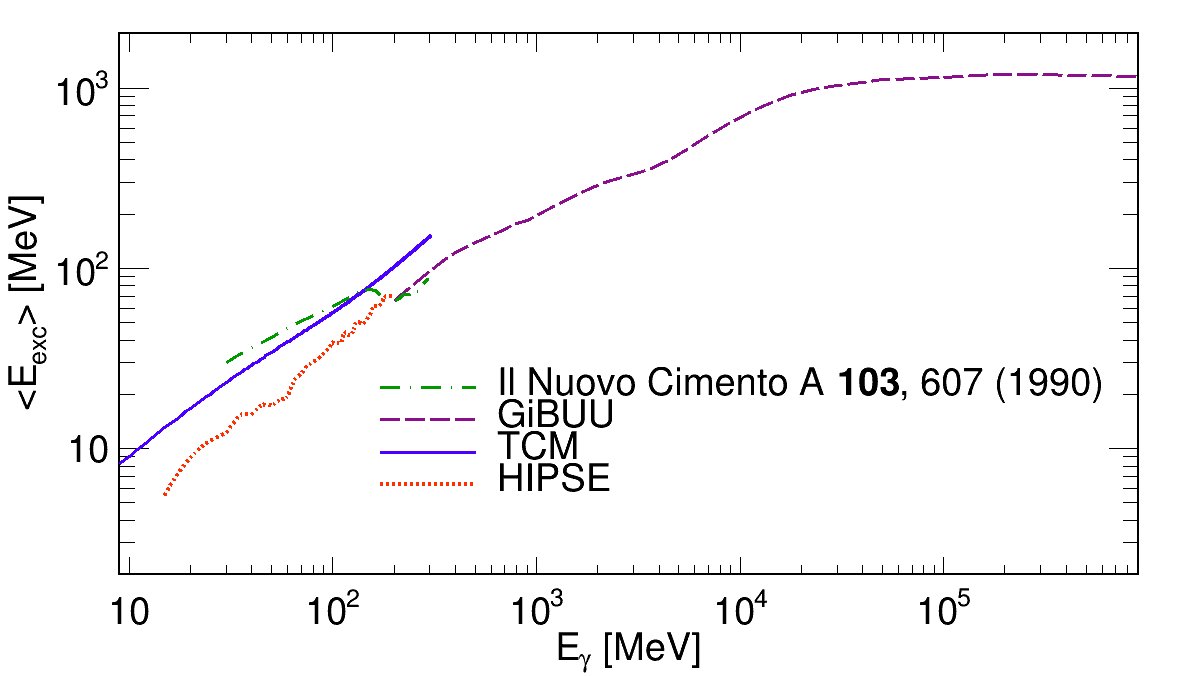}
\caption{Average excitation energy
as a function of energy of the photon in $\gamma+^{208}$Pb collisions in different models.}
\label{mean_energy}
\end{figure}

The comparison of the average excitation energy obtained with various
approaches for $^{208}$Pb bombarded by photons of various energies are displayed in Fig.~\ref{mean_energy}. The
experimental-like data comes from earlier calculation done with an intranuclear cascade model by Guaraldo et al.~\cite{guaraldo}
for several preactinides nuclei. The Two Components Model and HIPSE
estimation have been already discussed in Ref.~\cite{Jucha2025} 
but the GiBUU results anticipate 
the mean excitation energy for the photon-induced reaction also for higher photon energies up to $E_{\gamma}=10^6$~MeV. In the low-photon-energy range HIPSE and GiBUU give similar results although
the methods of estimation of excitation energy are rather different. It is due to the fact that both models (approximately) conserve the total energy of the system for reactions with $E_{\gamma}<$200~MeV.


\subsection{Production of nuclear remnants in $\gamma+^{208}$Pb reaction}
\begin{figure}[!bt]
\setlength{\unitlength}{0.1\textwidth}
\begin{picture}(4,7.9)
\put(-0.5,3.95){\includegraphics[width=3.30in]{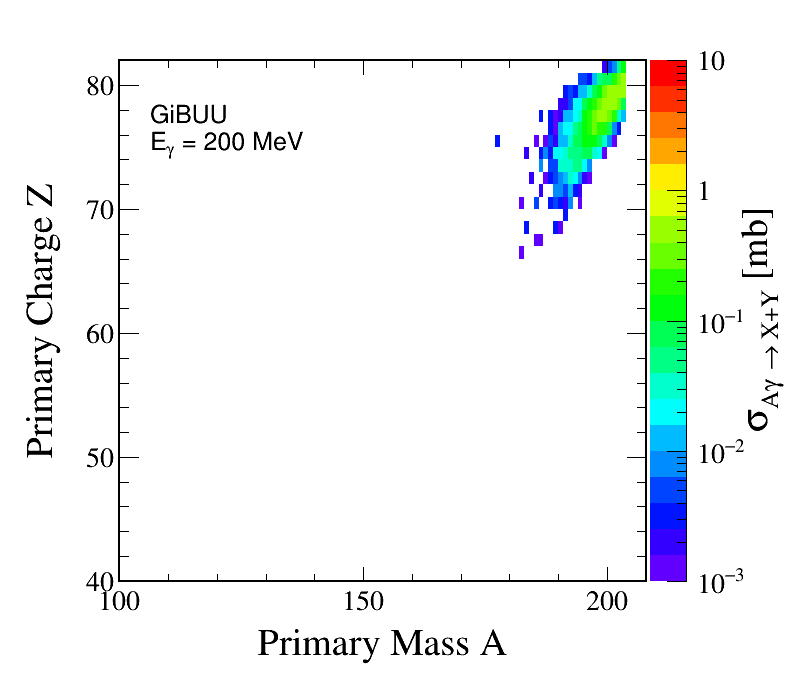}}
\put(-0.4,7.4){\bf \large a)}
\put(-0.5,0){\includegraphics[width=3.30in]{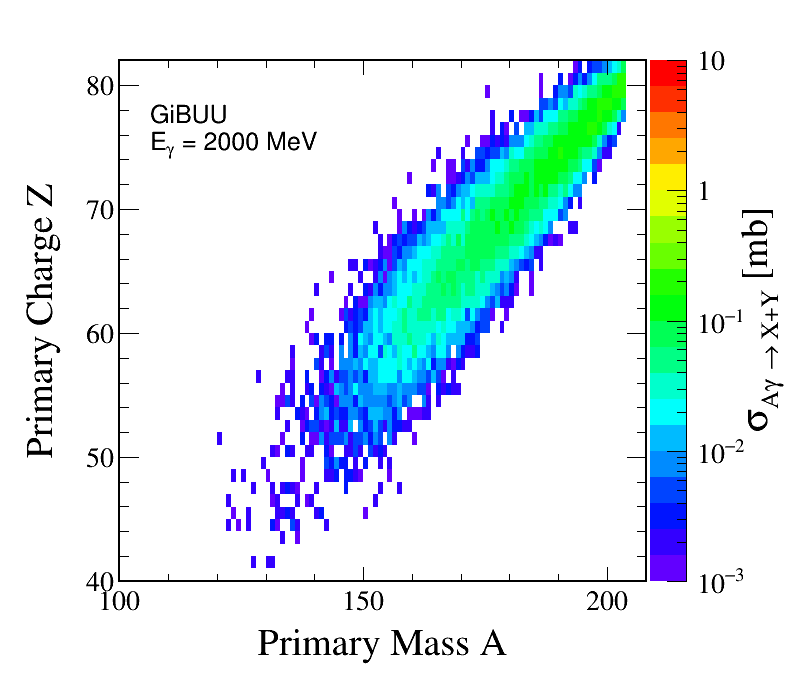}}
\put(-0.4,3.5){\bf \large b)} 
 \end{picture}
\caption{Cross section for production of primary fragments: GiBUU for collisions of 200~MeV and 2000~MeV photons in $\gamma+^{208}$Pb reaction. 
}
\label{pre_AZ}
\end{figure}

\begin{figure}[!bt]
\setlength{\unitlength}{0.1\textwidth}
\begin{picture}(4,7.9)
\put(-0.5,3.95){\includegraphics[width=3.30in]{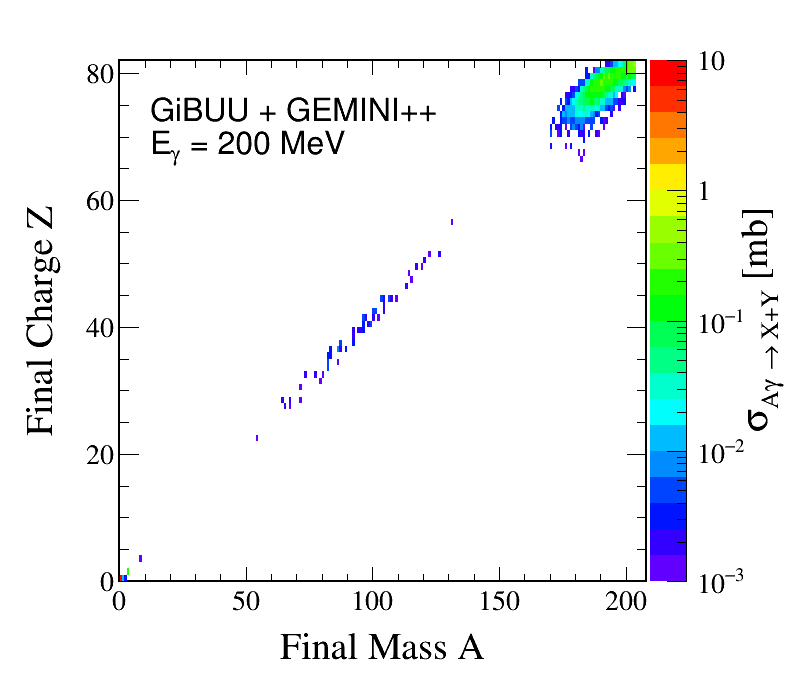}}
\put(-0.4,7.4){\bf \large a)}
\put(-0.5,0){\includegraphics[width=3.30in]{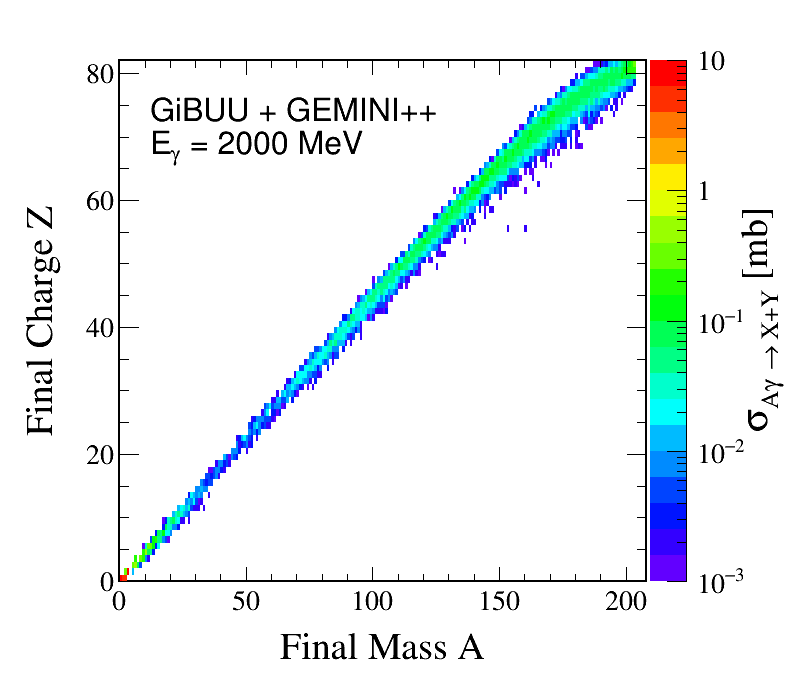}}
\put(-0.4,3.5){\bf \large b)} 
 \end{picture}
\caption{Cross section for production of final nuclei: GiBUU + GEMINI++ for  200~MeV and 2000~MeV photons in $\gamma+^{208}$Pb reactions. 
}
\label{post_AZ}
\end{figure}

The nuclei created in this study by GiBUU, are
highly excited. 
The ensemble of nuclei produced after the intranuclear cascade (Fig.~\ref{pre_AZ})
depends on the incident photon energy: the higher photon energy, the richer the
distribution of masses and charges of the residual nuclei are created. 
This trend is continued in Fig.~\ref{post_AZ}, where the final 
mass-charge distribution obtained after remnant de-excitation is displayed.  

During the intranuclear cascade and thermalization, except of nucleons, also heavier nuclear objects can be emitted. This is known from studies of nuclear spallation processes \cite{SMC_Pysz_2015,PISA_Budzanowski_Ni,PISA_Budzanowski_Ni_175,PISA_Fidelus_Al}. Unfortunately, in the GiBUU model, the formation and emission of nuclear clusters is disregarded. Thus, resulting particle multiplicities are, to some extent, biased by this model restriction.    

Final distribution of nuclei, when remnant deexctitation is simulated, are shown in Fig.~\ref{post_AZ}. In extreme case, very exotic nuclei can be generated up to the proton or neutron drip lines or even behind. The de-excitation of produced hot nuclei are estimated with GEMINI++ or GEM2. Figure~\ref{post_AZ} displays the final mass-charge distribution for two photon energies $E_{\gamma}$= 200~MeV and 2000~MeV. The evaporation residues are above mass 150, the fission fragments are in the mass range (50 - 150) and intermediate mass fragments are below mass 50.
The particle and $\gamma$ emission are allowed in every step of de-excitation. Here, apart from protons and neutrons, also $\alpha$, deuterium, tritium evaporation is taken into consideration.

\subsection{Proton and neutron multiplicities in $\gamma+^{208}$Pb reaction
}

The GiBUU approach allows only for emission of nucleons.
No $\alpha$ particles or other clusters are included.
This limitation results in a possible enhancement of the neutron and proton
multiplicities. 

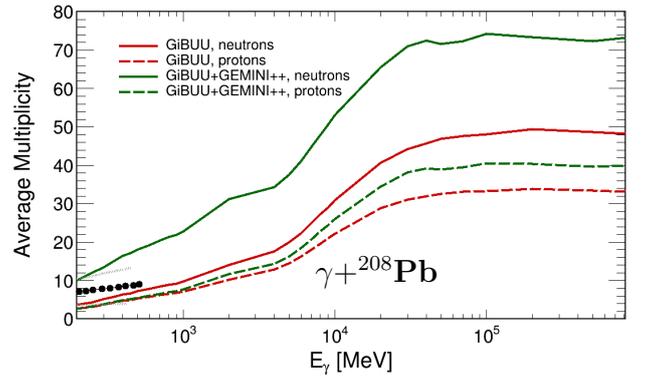
\begin{figure}[!bt]
\setlength{\unitlength}{0.1\textwidth}
\begin{picture}(4,5.5)
\put(-0.40,2.6){\includegraphics[width=3.50in]{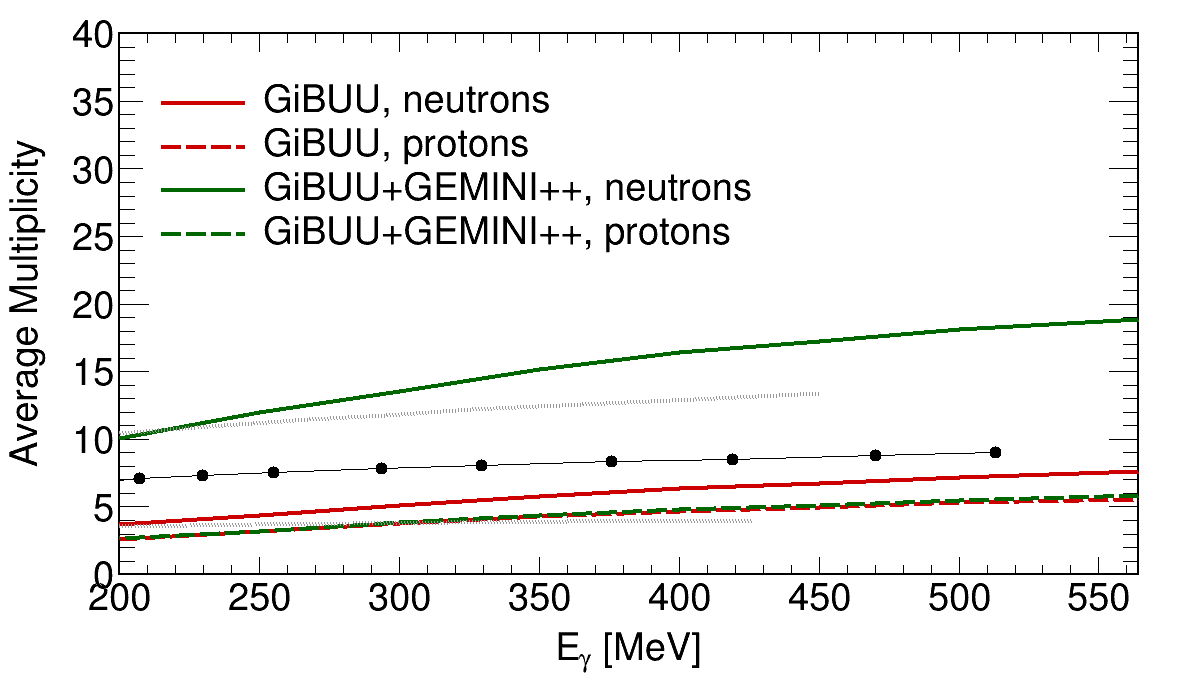}}
\put(-0.5,5.1){\bf \large a)}
\put(-0.40,0){\includegraphics[width=3.50in]{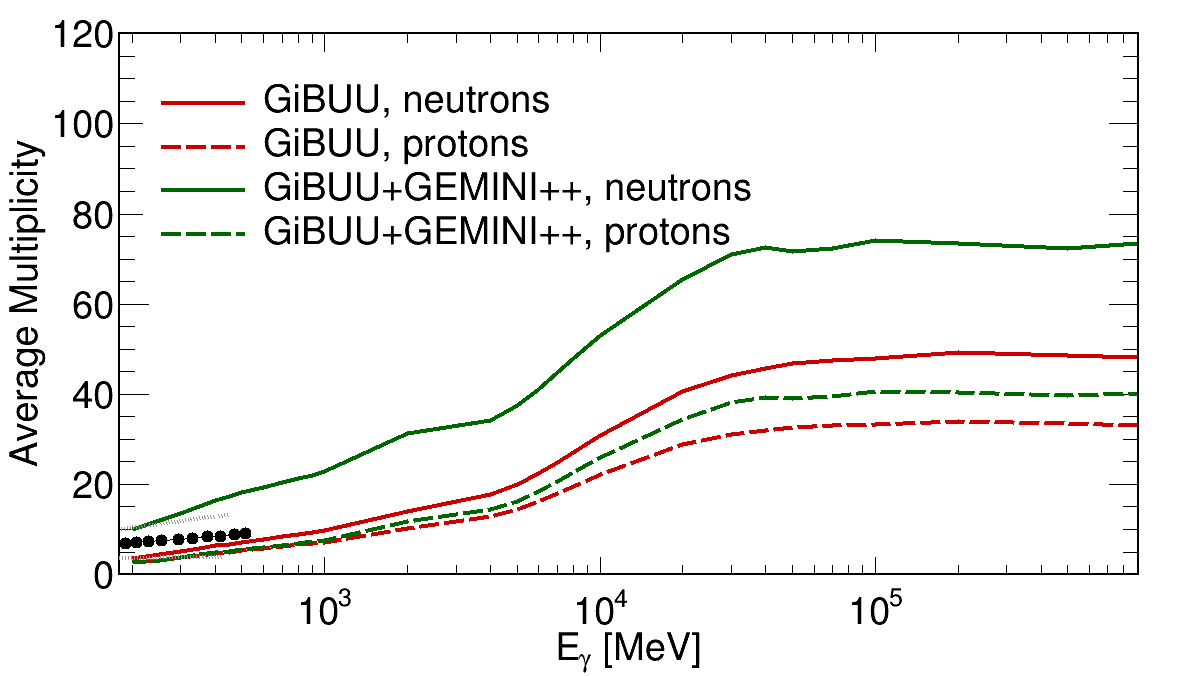}}
\put(3.,0.7){\bf \large $\gamma+^{208}$Pb}
\put(-0.5,2.5){\bf \large b)}
 \end{picture}
\caption{Average neutron multiplicity: pre-equilibrium obtained in GiBUU
  (red line), final results with GEMINI++ de-excitation code (green line); (a) zoomed on the photon energy $E_{\gamma}$=(200, 600) MeV and (b) for full energy range up to $E_{\gamma}$=$10^6$ MeV. 
The extrapolation of experimental data \cite{LEPRETRE1982, NooN}  
(black dots) and band of uncertainty (dotted lines) are shown. The full lines are for neutrons and the dashed ones are 
for protons.}
\label{mean_neutron}
\end{figure}

Experimental data for the average neutron multiplicities are available only up to $E_{\gamma}$=140~MeV \cite{LEPRETRE1978,LEPRETRE1981,LEPRETRE1982}.
Their extrapolation (taken from \cite{NooN}) are compared (Fig.~\ref{mean_neutron}) with the mean neutron number
emitted during the intranuclear cascade phase in GiBUU and the final results 
(GiBUU + GEMINI++). 
The multiplicities of the emitted particles depend not only on
the energy E$_{\gamma}$ of the photon, which excites the collision partner, but also on amount of energy stored in equilibration stage. 

The centroid of the neutron multiplicity distribution is moving 
towards the higher numbers when the de-excitation stage is included
(GiBUU+GEMINI++) in comparison to the primary emission (GiBUU).
The total multiplicities of neutrons coming from both stages seem to be too
large compared to the experimental data shown in Fig.~\ref{mean_neutron}. 
The experimental average neutron multiplicity were extracted for photon-induced neutron production from Pb in $E_{\gamma}=(10-140)$~MeV in Ref.~\cite{LEPRETRE1982} and later extrapolated in Ref.~\cite{NooN} to higher photon energies.

The multiplicity distribution of neutrons strongly depends
on excitation energy of residual nucleus. Lets consider the average neutron multiplicity for de-excitation process $\langle M_n^{sec.}\rangle$.
\begin{table}[!bt]
\caption{ The average multiplicity for secondary emission of neutrons $\langle M^{sec.}_n \rangle$ and protons  $\langle M^{sec.}_p \rangle$  for given excitation energy $E_{exc}$.
 (see Fig.~3 and 6 ).}
  \begin{center}
\begin{tabular}{cccc}
\hline &&&\\
 $E_{exc}$ (MeV) & \quad  $\langle E_{\gamma} \rangle$ (MeV) \quad  &  \quad $\langle M^{sec.}_n \rangle$  \quad & \quad  $\langle M^{sec.}_p \rangle$ \quad   \\ &&&\\\hline&&&\\
50  & 70   &   5     & 0             \\
150 & 200  &  13     & 0.14             \\
500 & 2000 &  25     & 12             \\
 &&&\\\hline
\end{tabular}
\end{center}
\label{tab:mp}
\end{table}

The combined information from Figs.~\ref{mean_energy} and \ref{mean_neutron} is presented in Table~\ref{tab:mp}, where the photon energy is translated into the excitation energy of the nucleus and the mean neutron multiplicity is estimated.

This average secondary neutron multiplicities $\langle M^{sec.}_n\rangle$ come only from the final stage of reaction where particles are emitted during deexcitation of the hot nucleus (see Fig.~\ref{mean_neutron}).
The situation for proton emission is less regular. 
Let's compare the red and green dashed lines in Fig.~\ref{mean_neutron}.
At $E_{\gamma}$ = 70~MeV there is almost no emission of protons.
For $E_{\gamma}$ = 200~MeV single proton is emitted with probability (number
of single protons per event) of around 0.14.
 At high excitation energy, $E_{\gamma}$ = 2000~MeV, emission of a bigger number of protons is possible.

\begin{figure}[!bt]
\setlength{\unitlength}{0.1\textwidth}
\begin{picture}(4,5.5)
\put(-0.50,2.55){\includegraphics[width=3.50in]{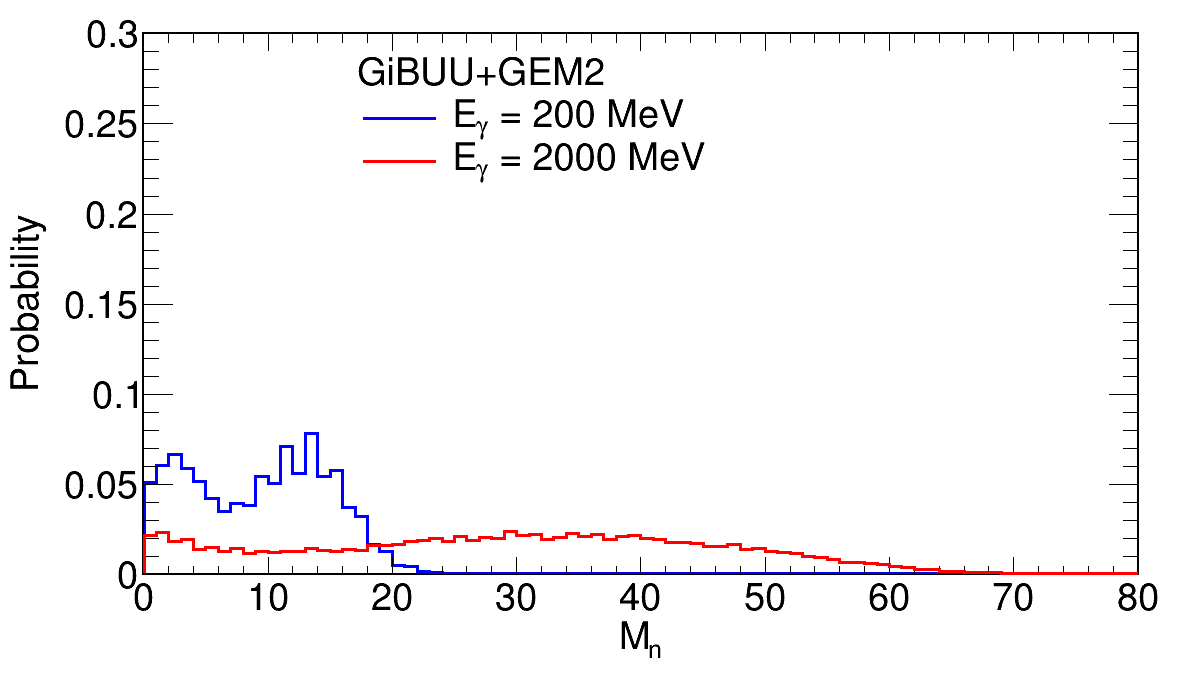}}
\put(2.5,4.9){\bf \large neutrons}
\put(-0.5,5.1){\bf \large a)}
\put(2.,3.9){\bf \large $\gamma+^{208}$Pb}
\put(-0.50,0){\includegraphics[width=3.50in]{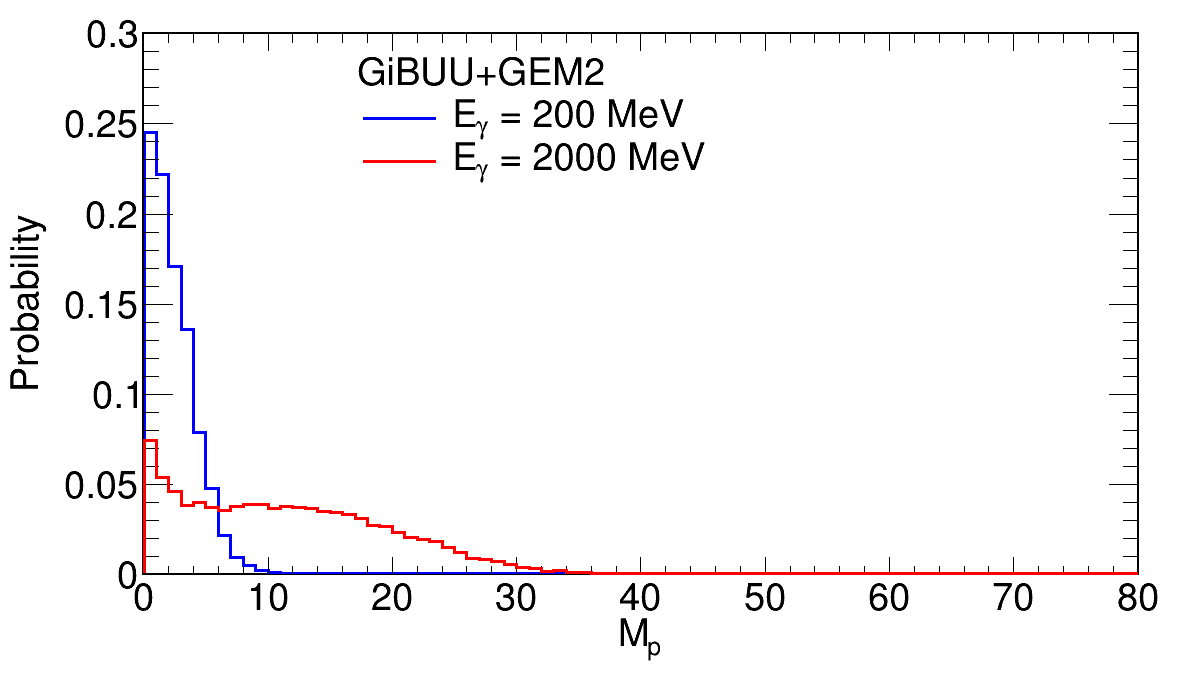}}
\put(2.5,2.3){\bf \large protons} 
\put(-0.5,2.5){\bf \large b)} 
\put(2.,1.5){\bf \large $\gamma+^{208}$Pb}
 \end{picture}
\caption{Neutron (a) and proton (b) multiplicities obtained with
  GiBUU+GEM2 for de-excitation of the initial nucleus at 
E$_{\gamma}$=200 (blue) and 2000 (red) MeV.}
\label{multiplicity}
\end{figure}
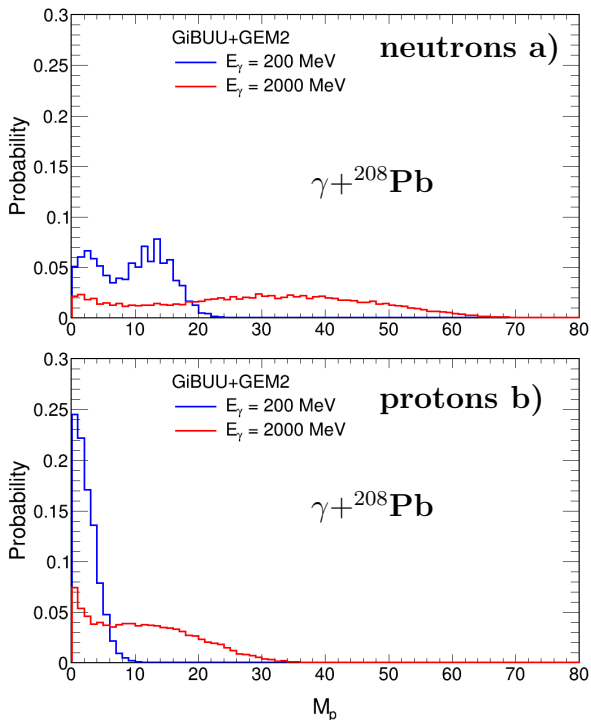
\begin{figure}[!bt]
\setlength{\unitlength}{0.1\textwidth}
\begin{picture}(4,5.6)
\put(-0.5,2.6){\includegraphics[width=3.50in]{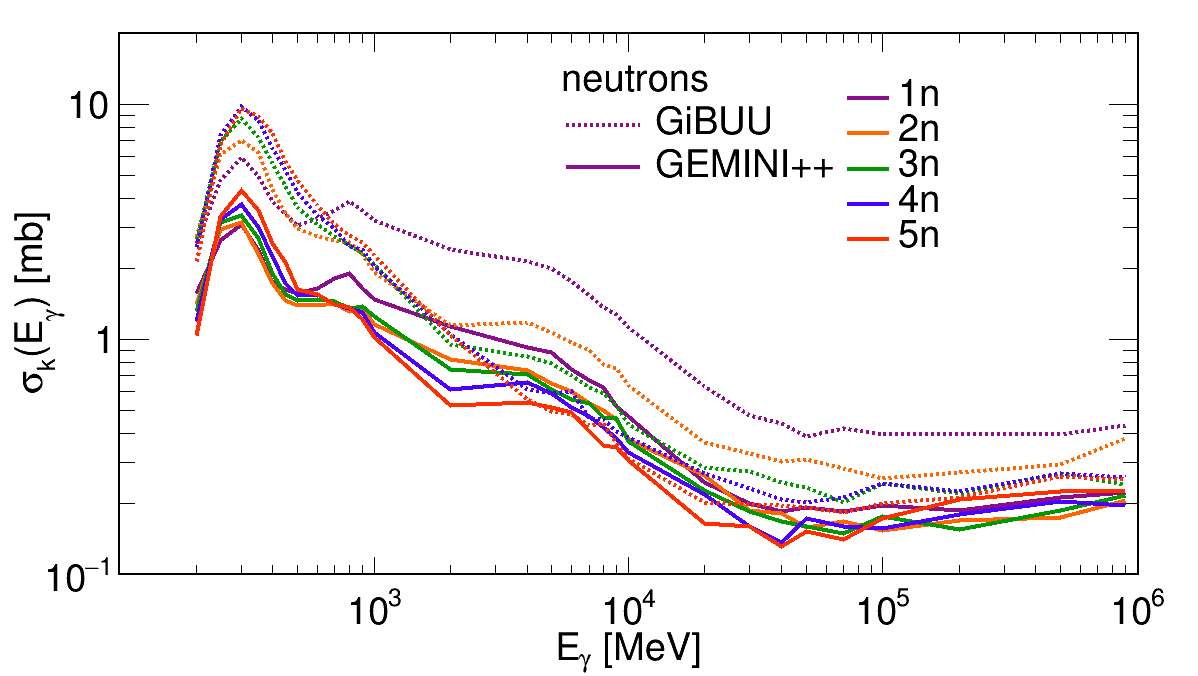}}
\put(-0.5,4.9){\bf \large a)}
\put(1.3,3.2){\bf \large $\gamma+^{208}$Pb}
\put(-0.5,0){\includegraphics[width=3.50in]{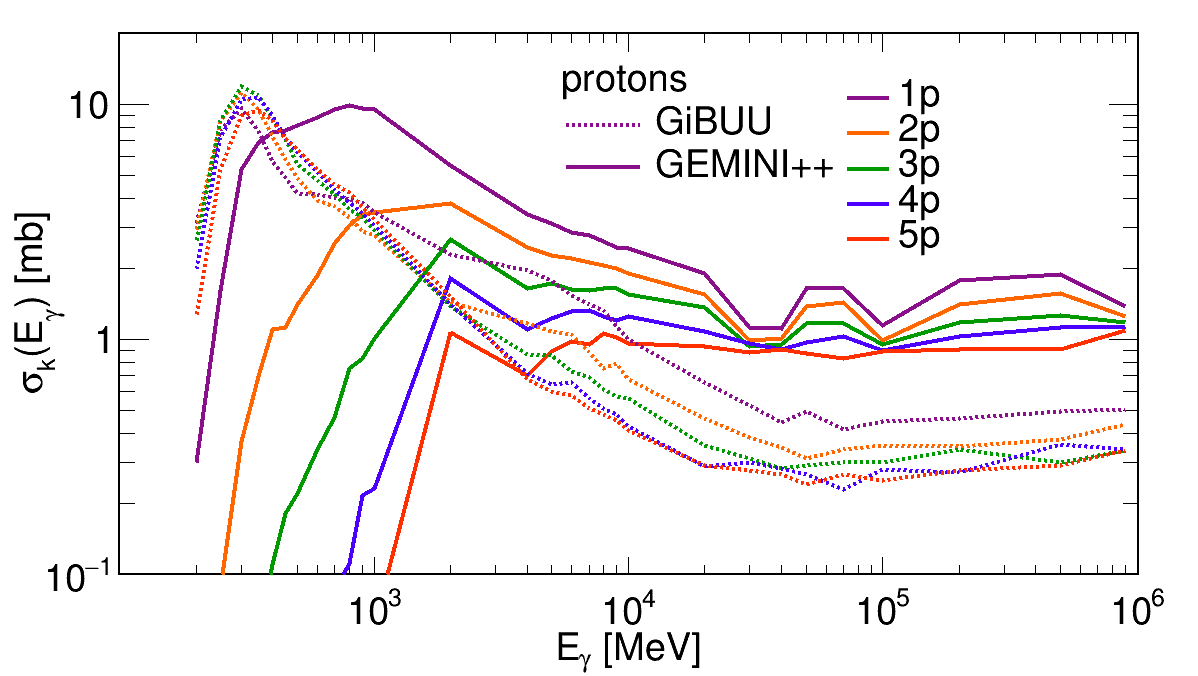}}
\put(-0.5,2.3){\bf \large b)} 
\put(1.3,0.56){\bf \large $\gamma+^{208}$Pb}
 \end{picture}
\caption{Neutron a) and proton (b) cross section for different multiplicities  for the emission of given number (k=1,2,3,4,5) of particles, obtained with
GiBUU + GEMINI++. }
\label{Pre-post_multi_n}
\end{figure}
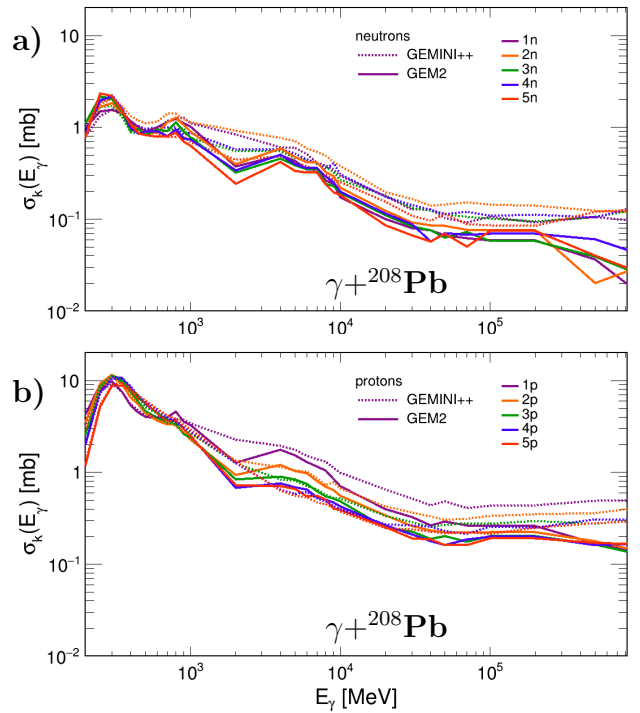
\begin{figure}[!bt]
\setlength{\unitlength}{0.1\textwidth}
\begin{picture}(4,5.6)
\put(-0.5,2.6){\includegraphics[width=3.50in]{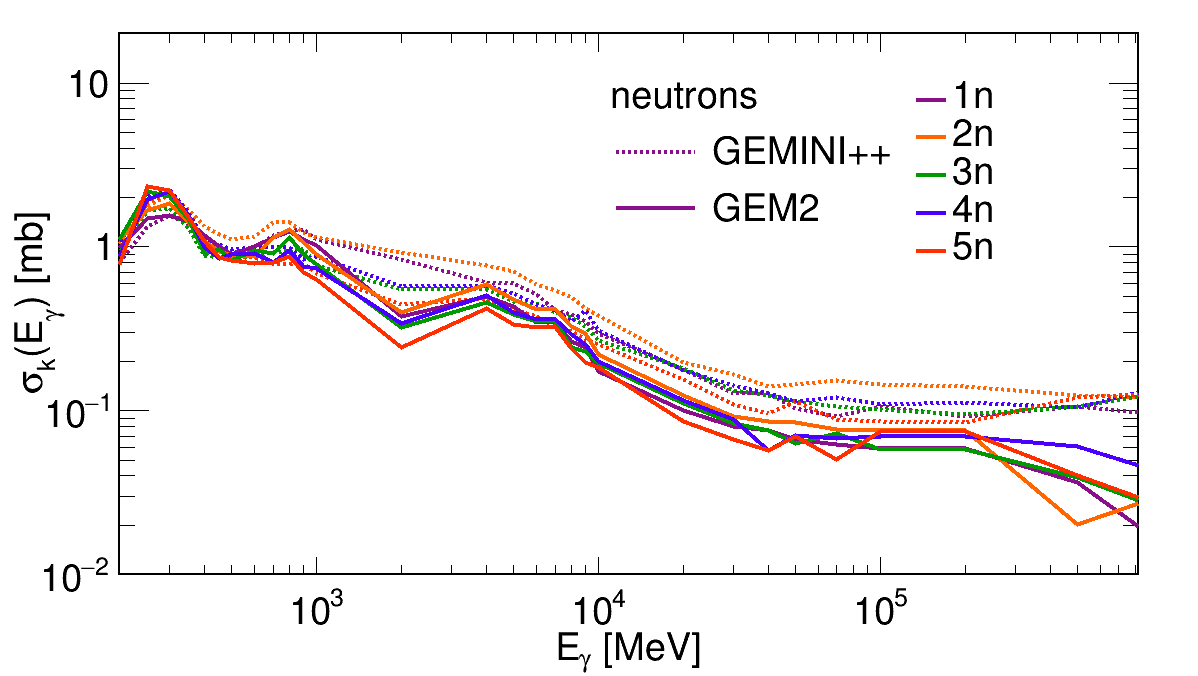}}
\put(-0.5,4.9){\bf \large a)}
\put(1.,3.2){\bf \large $\gamma+^{208}$Pb}
\put(-0.5,0){\includegraphics[width=3.50in]{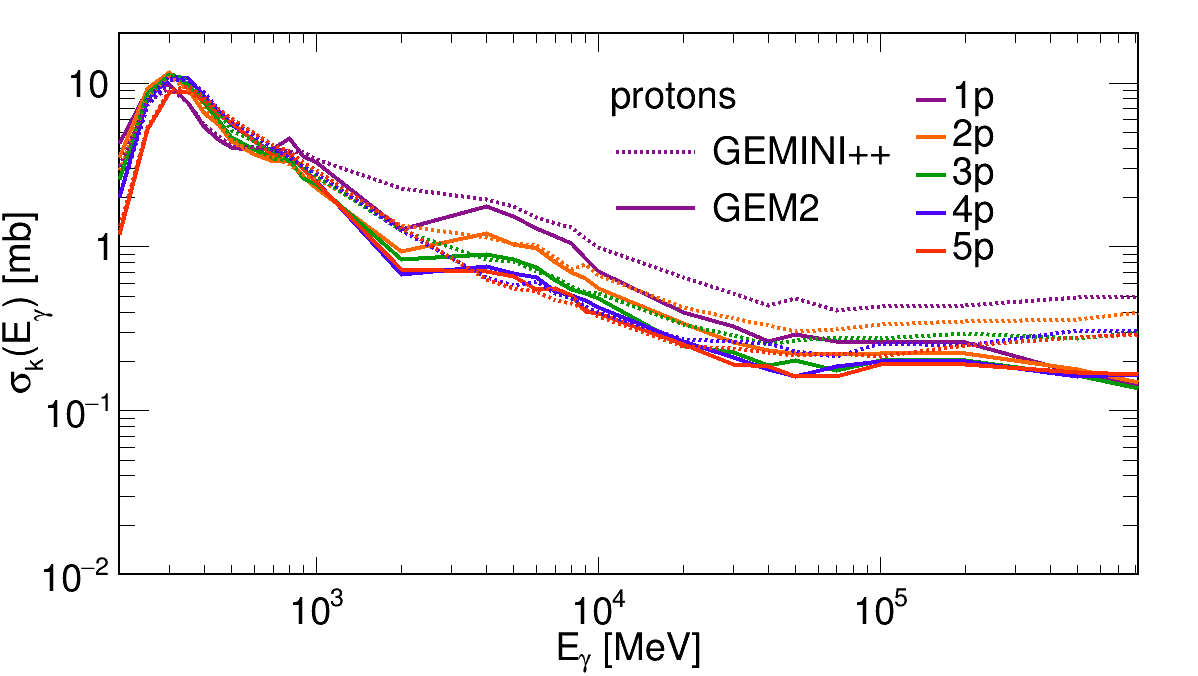}}
\put(-0.5,2.3){\bf \large b)} 
\put(1.,0.56){\bf \large $\gamma+^{208}$Pb}
 \end{picture}
\caption{Comparison of the GEMINI++ (dashed lines) and GEM2 (solid lines) de-excitation models cross section for a given number (k=1,2,3,4,5) of: a) neutrons, b) protons.}
\label{summary_emission}
\end{figure}

Figure~\ref{multiplicity} shows a comparison of multiplicity distributions of protons and 
neutrons emitted at two different photon energies: 
E$_{\gamma}$=200 (blue) and 2000 (red) MeV. They are calculated with combination of the GiBUU model for first step of reaction (collision phase) and GEM2 model for de-excitation of remnant nuclei. 
Large multiplicities of neutrons and protons are obtained,
especially at $E_{\gamma}$ = 2000 MeV.
For lower photon energies mainly single neutrons and protons are emitted. 

Increasing the energy E$_{\gamma}$ of photons reacting with $^{208}$Pb the distribution of the emitted particles is widening. In the pre-equilibrium stage calculated with GiBUU starting from 200~MeV, mostly few neutron events are predicted. 

Large number of protons that come from the GiBUU, are due to the fact that 
in pre-equilibrium stage only nucleons are considered, thus no $\alpha$,
deuterons, tritons, and $^{3}$He are taken into account.  

The cross section for different multiplicities as a function of photon energy $E_{\gamma}$ 
are presented separately for neutrons (Fig.~\ref{Pre-post_multi_n} a) and protons (Fig.~\ref{Pre-post_multi_n} b). 

In the first stage, a few particles are emitted and
the primary fragments have enough energy to release secondary
particles. Thus, the probability of emission of 1, 2 or 3 neutrons 
decreases. A similar tendency is visible in Fig.~\ref{Pre-post_multi_n}
(b) for the emission of protons.

Figs.~\ref{summary_emission} a) and \ref{summary_emission} b) compare the low multiplicity distributions for nucleons calculated with GiBUU and two different de-excitation codes: GEMINI++ (dashed line) and GEM2 (solid line). The results are very similar, especially for protons. For
neutrons the cross-sections for photon energy larger than
1 GeV are higher in GEMINI++ than in GEM2.

Although the basic results
behind these two models differ, the final results are very similar,
especially for protons. For neutrons the cross-section for photon energy
larger than 1000~MeV are somewhat higher in GEMINI++ than in GEM2.

\subsection{Integrated UPC cross sections}
The method, to obtain the cross section for the ultrarelativistic
ultraperipheral Pb+Pb reaction at $\sqrt{s_{NN}}=5.02$~ TeV was described
in detail in \cite{Jucha2025}. The main idea is the integration of the
electromagnetic flux with probabilities for a given number of neutrons or protons over a wide
photon energy range and the impact parameter. 
The following combinations of models are used to estimate the neutron or
proton emission or remnant production (presented also in Table~\ref{tab:mod}):\\
(1) The two-component model (TCM - Dirac
  delta + step-like probability function including GEMINI++) combined with GEMINI++ for E$_{\gamma}<$~200~MeV and constant probability for higher photon energies (TCG++);\\
(2) The two-component model with GEMINI++ for E$_{\gamma}<$~200~MeV and GiBUU with GEMINI++ for higher energies (TGG++);\\
(3) The two-component model with GEMINI++ for E$_{\gamma}<$~200~MeV and GiBUU with GEM2 for higher energies (TGG2);\\
(4) The HIPSE with GEMINI++ for E$_{\gamma}<$~200~MeV and GiBUU with GEMINI++ for higher energies (HGG++).\\
(5) The EMPIRE model that covers the photon energy up to 900~MeV.
\begin{table}[!bt]
\caption{ Different combinations of the models used in the present calculations.}
\begin{tabular}{cc|cc|l}
\hline \multicolumn{2}{c|}{}&\multicolumn{2}{c|}{}&\\
\multicolumn{2}{c|}{$E_{\gamma}<$200~MeV}&\multicolumn{2}{c|}{$E_{\gamma}>$200~MeV}&name\\
 \multicolumn{2}{c|}{}&\multicolumn{2}{c|}{}&\\
 \hline &&&&\\
$\gamma$+nucleus&deexcitation&$\gamma$+nucleus&deexcitation&\\
 &&&&\\
 \hline &&&&\\
TCM&GEMINI++&TCM&const.prob.& TCG++\\
TCM&GEMINI++&GiBUU&GEMINI++& TGG++\\
TCM&GEMINI++&GiBUU&GEM2& TGG2\\
HIPSE&GEMINI++&GiBUU&GEMINI++& HGG++\\
&&&&\\
\hline
\end{tabular}
\label{tab:mod}
\end{table}
Constant probability at large energies was introduced already in our previous paper [2]. Of course, this is a strong assumption but was inspired by the Saclay experimental data \cite{LEPRETRE1982} and used in \cite{NooN,RELDIS}. Long tails of probability in neutron and proton energies are probably due to preequilibrium processes as the ones considered in GiBUU and also in section IV of the present paper.

In Tab.~\ref{tab:tails} we compare results obtained with GiBUU
combined with two de-excitation models: GEMINI++ and GEM2 and with the TCM approach [Eq.(~\ref{two-component_model})] applied 
naively to proton emissions\footnote{The model was constructed
in \cite{Jucha2025} to understand neutron emissions.}. Here, only one photon exchange is taken into consideration (see Fig.~1 (a) from \cite{Jucha2025} ) as the multiple photon exchange were estimated to give contribution of the order of 10~$\%$. Similar results were presented in \cite{RELDIS}.
In comparison to TCM the calculation with the GiBUU model combined with 
de-excitation codes gives almost 2 times lower cross sections. 
Both GEMINI++ and GEM2 show a similar tendency as 
seen already in Fig.~\ref{summary_emission}.

\begin{table}[!bt]
\caption{Total cross sections (in barn) for a fixed number of neutron in UPC for reaction of $^{208}$Pb+$^{208}$Pb with energy $\sqrt{s_{NN}} = 5.02$~TeV.  
The integration is done only for energies E$_\gamma$ $>$ 200 MeV. The two-component model (TCM) with GEMINI++ and constant neutron probability equal to P(E$_{\gamma}$ = 200~MeV) - (TCG++) are compared to the GiBUU with GEMINI++ (TGG++) and the GiBUU with GEM2  (TGG2) results. Here, only one photon exchange 
is taken into consideration (see Fig.~1 (a) from \cite{Jucha2025} ).}
  \begin{center}
\begin{tabular}{lccc}
\hline &&&\\
   & \quad \quad TCG++ \quad \quad & \quad \quad TGG++ \quad \quad & \quad \quad TGG2 \quad \quad \\ &&&\\\hline&&&\\
1n & 2.294        & 1.308            & 1.160 \\
2n & 3.210        & 1.553            & 1.207 \\
3n & 2.513        & 1.160            & 1.155 \\
4n & 3.324        & 1.325            & 1.136 \\
5n & 2.940        & 1.156            & 1.079 \\&&& \\\hline
\end{tabular}
\end{center}
\label{tab:tails}
\end{table}


The total cross sections for a neutron, proton emission, and isotope production are compared for various
scenarios in Tab.~\ref{tab:final}.

\begin{table}[!bt]
\caption{Total cross sections (in barn) for a fixed number of neutrons and protons
  emissions and isotope production in UPC with energy $5.02$~TeV. 
  The cross section integration ingredients depends on the photon energies. For photon energies E$_\gamma <$~200~MeV available are: 
  the TCM, HIPSE and EMPIRE approaches. For energies E$_\gamma\geq$ 200~MeV, the constant probability (TCG++) and the GiBUU with GEMINI++ (TGG++) and GEM2 (TGG2); HIPSE with GiBUU and GEMINI++ (HGG++) as well as EMPIRE results are presented. Here, only one photon exchange is taken into consideration (Fig.~1 (a) from \cite{Jucha2025} ). }
\begin{tabular}{lcccccc}
\hline &&&&&&\\
 $E_{\gamma}$& TCG++  & TGG++ & TGG2  & HGG++& EMPIRE&ALICE \cite{ALICE502a,ALICE_protons} \\ 
 &&&&&&\\
 \hline
&&&&&&\\
1n & 98.79    & 97.37& 97.23  &113.21 &98.90 & 108.4$\pm$3.90   \\
2n & 25.31    & 23.24& 22.90  &14.34  &23.39 &25.0$\pm$1.30     \\
3n & 6.03      & 4.24& 4.23   &4.24   &3.91  &7.95$\pm$0.25     \\
4n & 6.32      & 3.69& 3.51   &3.41   &2.28  &5.65$\pm$0.33     \\
5n & 4.91      & 2.56& 2.49   &2.79   &1.05  & 4.54$\pm$0.44    \\ &&&&&&\\
\hline&&&&&&\\
1p & 6.59  & 12.54&  13.18   & 5.25   & 2.36 & 40.4$\pm$1.6   \\
2p & 0.44   & 6.63&  3.71    & 4.74   &0.01  & 16.8$\pm$3.7   \\
3p & 0.01   & 3.40&  2.35    & 4.60   &0.01  & 6.8$\pm$2.2     \\ &&&&&&\\
\hline&&&&&&\\
Pb & 150.47  & 125.12&124.87  &  130.7 &129.51   & 157.5$\pm$4.6   \\
Tl & 5.86    & 5.42&4.97    & 7.24   & 0.001   &  40.4$\pm$1.6   \\
Hg & 9.67    & 4.88&4.59    & 4.92   &0.0007   &  16.8$\pm$3.7   \\
Au & 2.43    & 4.29&3.97    & 4.25   & 0.00001 &   6.8$\pm$2.2   \\ &&&&&&\\
\hline
\end{tabular}
\label{tab:final}
\end{table}
Both TCM and GiBUU combined with GEMINI++ or GEM2
badly fail to describe the large ALICE cross section \cite{ALICE_protons} for the one-proton emission ($\sigma_{1p}$ $\approx$ 40 b).
While for neutron channels all models give right order of magnitude, the situation for proton channels is very different. Here, all models underestimate the new ALICE proton data \cite{ALICE_protons}. The situation with isotope production is similar. 

\begin{figure}[!bt]
\includegraphics[width=8cm]{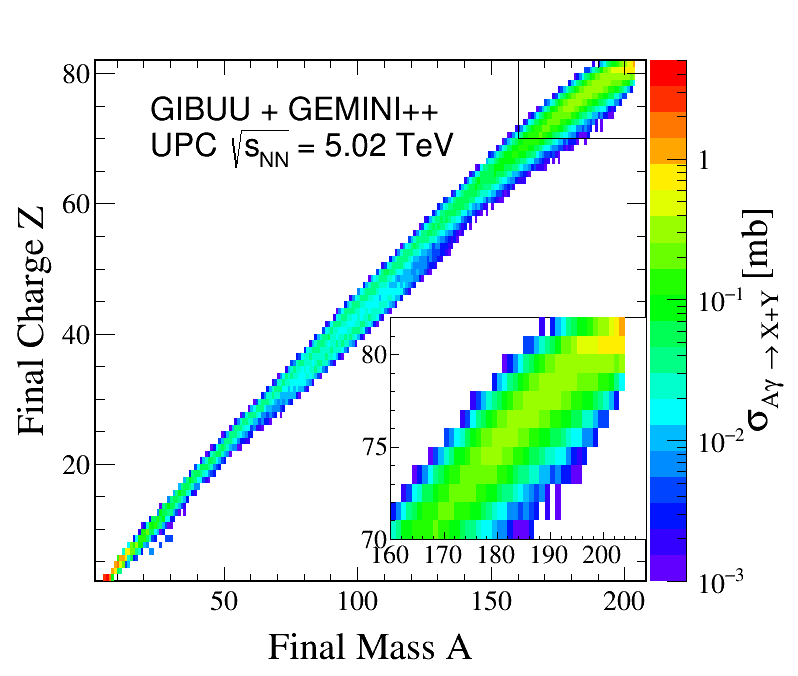}
\caption{Cross section for production of nucleus remnants in $^{208}$Pb+$^{208}$Pb UPC. 
The inset plot is the zoom on mass range A=(160-208).}
\label{UPC_AZ}
\end{figure}

In Fig. \ref{UPC_AZ} presented is a cross section calculated with the combination of GiBUU and GEMINI++ models. As expected and measured in \cite{ALICE_protons} the  dominant yield is foreseen for target-like nuclei close to A=208 and Z = 82.  According to our predictions, except the lead nuclei, in ultraperipheral collisions the final de-excited nuclei of mass A $>$ 170 and charge Z $>$ 70 are mostly populated. 

Comparing Tables~\ref{tab:tails} and \ref{tab:final} it is clear that most of the contribution from the energy region $E_{\gamma}<$~200~MeV as there the absorption cross section is large (see Fig.~\ref{Photo_abs}). 

\section{\label{Sect.IV} On the pre-equilibrium production of one
 proton channels above giant dipole resonance}

The $\gamma + ^{208}$Pb$ \to$ hadrons is the main channel of the proton/neutron production in $^{208}$Pb + $^{208}$Pb 
UPC collisions.
The emission of nucleons in scattering photons off nuclei
was studied some time ago for $^{12}$C$(\gamma,pn)$ and
$^{12}$C$(\gamma,pp)$ \cite{Watts2000} using tagged photons at MAMI 
with $E_{\gamma}$~=~150-700~MeV and more recently
in a PhD thesis of R. Williams \cite{Williams2024} also 
for $\gamma + ^{12}$C collisions at JLab. In the latter case
the emphasis was on multiproton production. 
To our knowledge, no experimental input
for our $\gamma + ^{208}$Pb reaction exists.
Therefore, in the following we have to rely on models.

 As shown in Table~\ref{tab:final} the cross section for one-proton ($1p$) emission
($1p$ $\equiv$ $1px$, where $x$ means production of anything but not protons) 
in the statistical codes is very small compared to the huge cross 
section of about 40~barns measured by the ALICE collaboration 
\cite{ALICE_protons}. 

At low energies photons interact with the whole nucleus 
and excite giant dipole resonance (GDR).
At higher energies photons start to interact with individual nucleons
inside the nucleus.
The interaction of photons with nucleons is complicated and strongly energy dependent
as different mechanisms play the role for $E_{\gamma} >$ 150 MeV.
Only a few simple final states were carefully measured
at Bonn Elsa, MAMI at Mainz or at JLab. Some simple final
states are included in intra-nuclear cascade codes such as
PICA \cite{PICA95} or RELDIS \cite{RELDIS}.
We will show below that even this complicated and elaborated treatment
may be not sufficient.
We wish to note here that even for $\gamma + d \to p + n$ process
the mechanism of reaction is not well understood \cite{d_JLAB}.

At even larger photon energies, say $E_{\gamma} >$ 1~GeV,
one is not able to control all final states and one is forced
to concentrate rather on inclusive production of protons and neutrons.
This high-energy region of the $\gamma N$ interaction and its
impact on the $\gamma A$ interaction was not well documented
in the literature.

In this section we discuss how the physics of $\gamma A$
interactions changes when increasing gradually photon energy.
Intentionally, restricted by available data, we will focus on production of one-proton and one-neutron only
in processes initiated by collisions of photon with individual
nucleons.
The large cross section for one-proton emission measured recently by
the ALICE collaboration is particularly interesting as no
existing model can explain it.

In this section we will try to estimate the maximally available
cross section for the one-proton emission.

Our total photoabsorption cross section is a sum of four different
components \cite{Klusek-Gawenda:2013ema}:\\
1) giant dipole resonance, $\sigma_{\gamma A}^{GDR}(\omega)$,\\
2) quasi deuteron mechanism, $\sigma_{\gamma A}^{QD}(\omega) $,\\
3) nucleon resonance region,\footnote{The ``resonance'' region includes also
  non-resonance contributions.} $\sigma_{\gamma A}^{res}(\omega) $,\\
4) partonic region, $\sigma_{\gamma A}^{part}(\omega)$
\begin{eqnarray}
\sigma_{\gamma A}(\omega)
= \sigma_{\gamma A}^{GDR}(\omega)
+ \sigma_{\gamma A}^{QD}(\omega)   
+ \sigma_{\gamma A}^{res}(\omega)
+ \sigma_{\gamma A}^{part}(\omega),
\label{photoabsorption_cross_section}
\end{eqnarray}
where $\omega$ is photon energy. As will be discussed in the following, for lowest photon
energies (GDR), practically no protons are emitted.
At higher photon energies, photon interacts rather with
individual nucleons and not with the whole nucleus.
While the produced neutron can be easily emitted, 
the proton experiences a Coulomb barrier. 
The Woods-Saxon single particle potential for neutrons and protons (including Coulomb part) are 
illustrated in Fig.~{\ref{fig:single_particle_potentials} as a function of distance from the center-of-mass of $^{208}$Pb nucleus.
The parameters of the potentials were tested in several papers
\cite{DW1978,DMSW1979,DNW1980,DSW1981,CDNSW1987}.
Proton emission becomes easier when $E_p > E_{barrier} \approx$ 10~MeV, which seems to have smaller influence for
$E_\gamma$~$>$~200~MeV.

\begin{figure}
\begin{center}
\includegraphics[width=7.5cm]{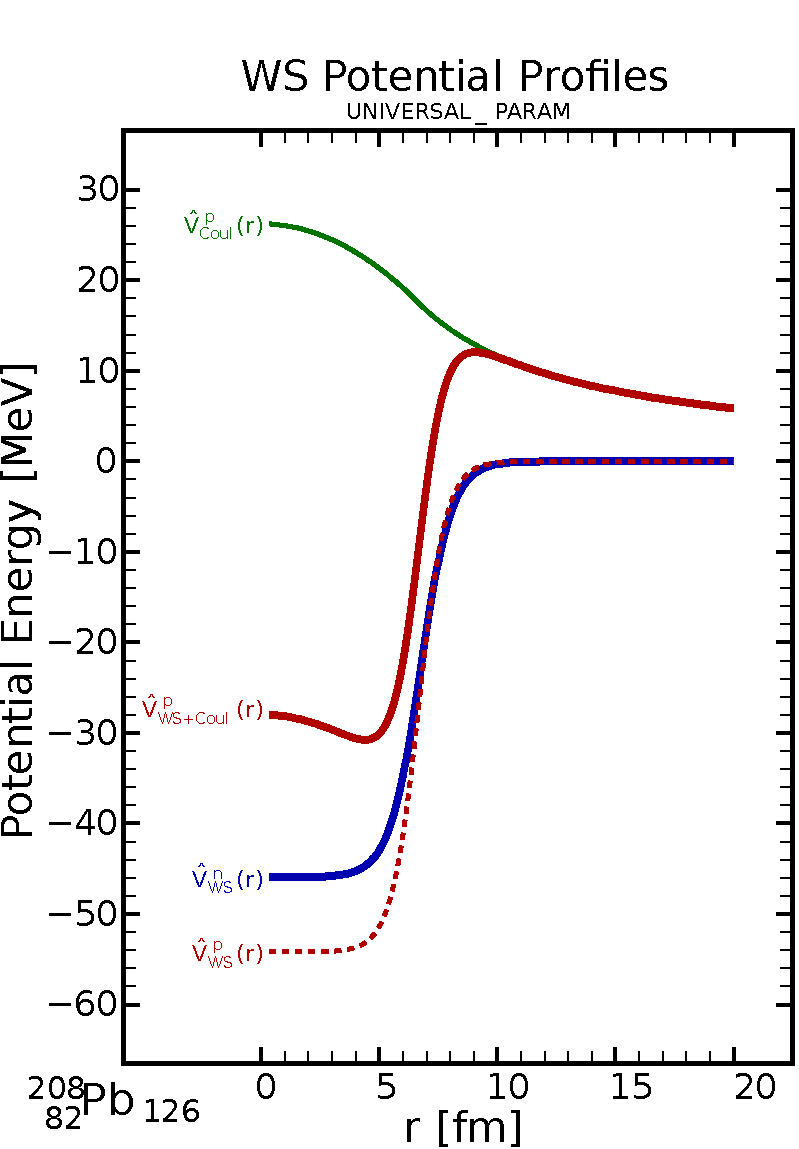}
\caption{The Woods-Saxon single particle potential for neutrons (blue
  line) $\hat{V}_{WS}^{n}$ and protons (dashed red line) $\hat{V}_{WS}^{p}$, Coulomb part $\hat{V}_{WS+Coul}^{p}$ (green line) and their sum (solid red lines) as a function of distance from 
center-of-mass of $^{208}$Pb nucleus.}
\label{fig:single_particle_potentials}
\end{center}
\end{figure}

It is obvious that the cross section for $1p$ emission must be smaller 
than the photoabsorption cross section for any $\omega$.
Then the absolutely maximal estimation of the one-proton cross section in UPC
can be approximated as:
\begin{eqnarray}
max\{\sigma_{AA \to 1p}\}
&=& max \{ \sigma_{AA \to p}^{QD} \}
+ max \{ \sigma_{AA \to p}^{res} \} \nonumber\\
&+& max \{ \sigma_{AA \to p}^{part} \}  \, ,
\label{decomposition}
\end{eqnarray}
where cross sections for one-side $1p$ emission (forward or backward) 
in UPC can be obtained as
\begin{eqnarray}
max\{ \sigma_{AA \to 1p}^{QD} \} &\approx& 
\int N(\omega, b) \sigma_{\gamma A}^{QD}(\omega) \; d^2b d\omega 
\; , \nonumber \\
max\{ \sigma_{AA \to 1p}^{res} \} &\approx&
\int N(\omega, b) \sigma_{\gamma A}^{res}(\omega) \; d^2b d\omega
\; , \nonumber \\
max\{ \sigma_{AA \to 1p}^{part} \} &\approx&
\int N(\omega, b) \sigma_{\gamma A}^{part}(\omega) \; d^2b d\omega
\; .
\end{eqnarray}
Similar equation can be written for one-neutron pre-equilibrium emissions.
If we apply parametrizations of different $\gamma Pb$ photoabsorption
components in (\ref{decomposition}) used in \cite{Klusek-Gawenda:2013ema} we get:\\
\begin{eqnarray}
max \{ \sigma_{AA \to p}^{QD} \} = 12.62 \; b, \nonumber \\ 
max \{ \sigma_{AA \to p}^{res} \} = 39.31 \;  b, \nonumber \\   
max \{ \sigma_{AA \to p}^{part} \} = 34.29 \;  b.
\end{eqnarray}
Adding maximal values for each component we get:
max$\left( \sigma_{AA \to 1p} \right)$ = 85 b
which is only about two times bigger than the cross section 
measured by the ALICE collaboration.

The estimation above is an absolute upper limit.
Not only protons but also neutrons can be produced in $\gamma+N$
collisions. The cross section with proton or neutron emission should not exceed the photoabsorption cross section. This must be so for each component (i), included in modeling the absorption cross section.
Then for different reaction mechanisms (i) the following simple
sum rule must be fulfilled:
\begin{eqnarray}
\sigma_{\gamma A}^{(i)}(\omega) >
&&Z \left( \sigma_{\gamma p \to pX}^{(i)}(\omega)
       + \sigma_{\gamma p \to nX}^{(i)}(\omega) \right)
  +\nonumber\\
&&N \left( \sigma_{\gamma n \to nX}^{(i)}(\omega) 
       + \sigma_{\gamma n \to pX}^{(i)}(\omega) \right)\; ,
\label{sum_rule}
\end{eqnarray}
where $i$ numerates individual processes.
Since for the different distinct mechanisms the relative production
of protons and neutrons is approximately known one can set upper limit
on {\bf $\sigma^{(i)}_{AA \to p}$}. This estimation includes production of
nucleon resonances and their decay.

The maximal value of the cross section can be therefore lowered.
This requires more detailed analyses.
We shall discuss all three contributions one by one.

We start from quasi-deuteron contributions.
We assume the simplest form:
\begin{equation}
\sigma_{\gamma A \to 1p}^{QD}(\omega) = C_p \times 
\sigma_{\gamma A}^{QD}(\omega) \; 
\label{QD}
\end{equation}
and fit $C_p$ to the existing data \cite{bremsstrahlung_Heidelberg} 
(see Fig.~\ref{fig:quasi_deuteron}). The giant dipole resonance, quasideuteron, nucleon
resonances and partonic components are shown in Fig.~\ref{fig:quasi_deuteron}. The solid red line corresponds to the QD absorption cross section. However, the proton emission cross section may be smaller than QD component of photoabsorption cross section due to the Coulomb barrier. We rescale the QD absorption cross section to the measured proton emission data.
We find $C_p \approx$ 0.3. Then
$\sigma_{AA \to 1p}^{QD} \approx$ 3.5 b, much less than 
the ALICE result.

\begin{figure}[!bt]
\setlength{\unitlength}{0.1\textwidth}
\begin{picture}(4,3)
\put(-0.5,0){\includegraphics[width=3.50in]{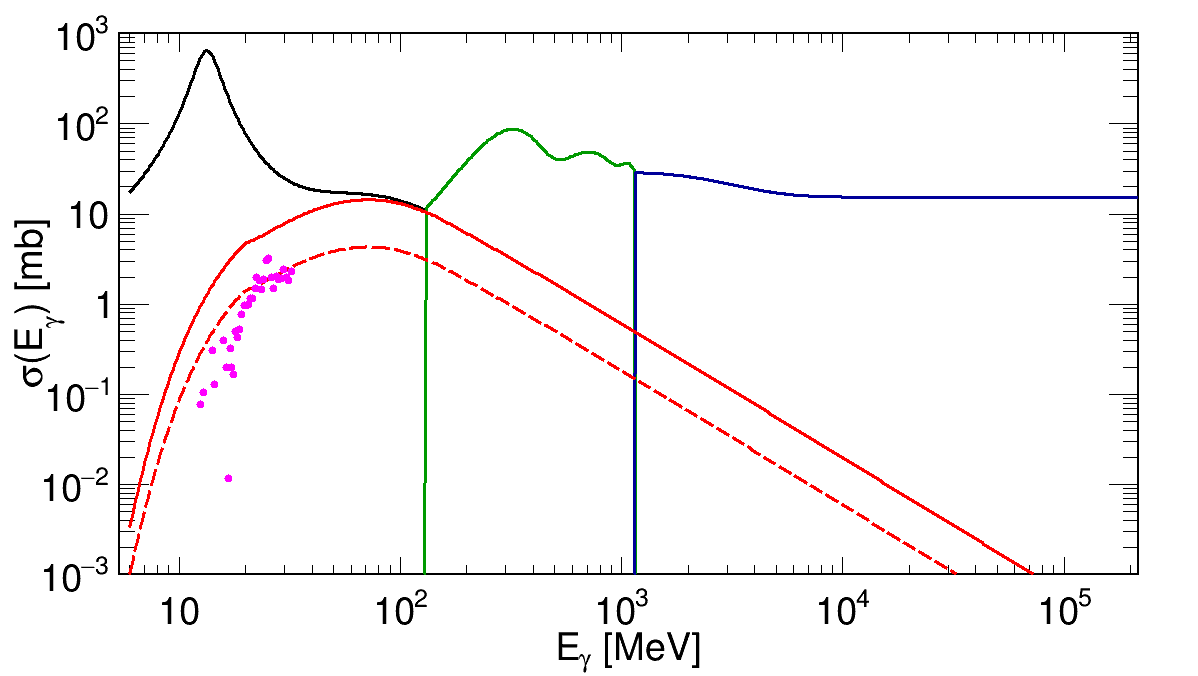}}
\put(0.7,2.45){\bf GDR} 
\put(0.9,1.5){\bf QD} 
\put(1.5,2.3){\bf nucl. res.} 
\put(2.7,1.8){\bf partonic} 
 \end{picture}
\caption{Cross section for $\gamma + ^{208}$Pb$ \to p$ and our
simple fit result. The giant dipole resonance, quasideuteron,
nucleon resonances and partonic components
(Eq.~\ref{photoabsorption_cross_section}) 
are shown. The experimental points \cite{DAHMEN1971140} are marked by solid circles. The red dashed line is reduced cross section (see text) in comparison to the red solid line.}
\label{fig:quasi_deuteron}
\end{figure}

Therefore, we also consider the resonance region. A representative 
example of the reaction to be considered are:
\begin{eqnarray}
\gamma + p &\to& \Delta^+ \to p \pi^0, \; 
P = {\Large\vert\langle}\frac{3}{2}, \frac{1}{2} \vert\frac{1}{2}, \frac{1}{2}, 1 ,0{\Large\rangle\vert}^2
= \frac{2}{3} \; ,     \nonumber \\
\gamma + p &\to& \Delta^+ \to n \pi^+, \; 
P = {\Large\vert\langle}\frac{3}{2},  \frac{1}{2} \vert \frac{1}{2}, -\frac{1}{2}, 1, 1{\Large\rangle\vert}^2
= \frac{1}{3} \; ,     \nonumber \\
\gamma + n &\to& \Delta^0 \to n \pi^0, \; 
P = {\Large\vert\langle}\frac{3}{2}, -\frac{1}{2} \vert \frac{1}{2}, -\frac{1}{2}, 1, 0{\Large\rangle\vert}^2
= \frac{2}{3} \; ,     \nonumber \\
\gamma + n &\to& \Delta^0 \to p \pi^-, \; 
P = {\Large\vert\langle}\frac{3}{2}, -\frac{1}{2} \vert \frac{1}{2}, \frac{1}{2}, 1, -1{\Large\rangle\vert}^2
= \frac{1}{3} \; .     \nonumber\\  
\label{Delta_excitations}
\end{eqnarray}
Different $\Delta$ isobars have different branching fractions. $P$ in Eq.(12) is probability of decay of $\Delta$ isobar to one of the $\pi N$ channels. The relative probability is calculated in terms of the Clebsch-Gordan coefficients.
There are other nucleon resonances (not visible in $\gamma A$ 
collisions) and a few continuum contributions
like:
\begin{eqnarray}
\gamma + p &\to& \pi^0 + p, \nonumber \\
\gamma + p &\to& \pi^+ + n, \nonumber \\
\gamma + n &\to& \pi^0 + n, \nonumber \\
\gamma + n &\to& \pi^- + p,
\end{eqnarray}
which start at $\omega >$ 180 MeV. The $\gamma N \to \pi N$
reactions can be treated microscopically including
meson-exchange currents and channel coupling \cite{SL1996}.
This treatment  starts to be almost impossible already
for $\gamma N \to \pi \pi N$ reactions.
Therefore microscopic treatment is limited only to some
energy windows. 

Another strongly populated final state is exclusive vector meson 
production $V$:
\begin{eqnarray}
\gamma + p &\to& V + p,    \nonumber \\
\gamma + n &\to& V + n.
\label{vector_mesons}
\end{eqnarray}
where $V = \rho^0, \omega, \phi, J/\psi$ 
\footnote{The $p/n$ proportions there are different than
for $\Delta$ resonances discussed  above.}.
The $\gamma N \to V N$ reactions can be successfully calculated either
within Regge approach \cite{V_Regge} 
(see also \cite{BMNSS2014} within tensor-pomeron model) 
or dipole model approach
\cite{V_dipole}. However, the $\gamma N \to V N^*$ processes
were not carefully studied within microscopic models,
but they also lead to sizable production of protons or neutrons.

The cross section for $\gamma+p \to \rho^0 + p$ only weakly depends
on energy and is 10-20~$\mu$b (see e.g.\cite{BMNSS2014}). 
At sufficiently high energy the cross section for $\gamma A \to \rho^0$ can be roughly estimated as $A \sigma(\gamma p \to \rho^0 p)$. Then the corresponding contribution to  $\gamma+A \to p$ is therefore 
about 0.8-1.6~mb, which constitutes sizable fraction of 
the $\gamma+A$ absorptive cross section (see Fig.~\ref{photoabsorption_cross_section}).
This inelastic cross section is of the same order of magnitude
as $\gamma + A \to \rho^0 + A$ \cite{FGSZ2016}.

More processes were discussed, e.g. in \cite{PICA95,P2011}.
Our estimation here has advantage that it exhausts by construction
the absorption cross section for processes on individual nucleons.
For the sake of simplicity, assuming that the $\Delta$ resonances
are representative for the whole ``resonance'' region,
we write:
\begin{eqnarray}
&&\sigma_{AA \to 1p}^{res} \approx 
\left( \frac{2}{3} \frac{Z}{A} + \frac{1}{3}\frac{N}{A} \right)
\sigma_{AA}^{res}  \; ,
\nonumber \\
&&\sigma_{AA \to 1n}^{res} \approx 
\left( \frac{1}{3} \frac{Z}{A} + \frac{2}{3}\frac{N}{A} \right)
\sigma_{AA}^{res}  \; .
\label{resonance_region}
\end{eqnarray}
We can see that even including the quasi-deuteron and resonance 
regions we are not able to understand the ALICE result for one proton 
emission (Tab.~\ref{tab:qd1}).

Therefore we consider also the highest energy, ``partonic'',
component. The production of protons or neutrons is subjected to
the mechanism of nucleon remnant fragmentation
(see e.g.\cite{LEPTO}).
The HERA data on leading neutron and proton production
in $\gamma^* p$ collisions showed that we did not
fully understand the underlying physics before the HERA results.
New mechanisms were proposed in \cite{HLNSS1994} and \cite{SNS1998}.
Combining the conventional at that time and ``new'' mechanisms 
requires a hybrid approach for fixed target experiments 
(see e.g.\cite{SBD1995}).
Since in the current paper we are interested just in proton and neutron
production, we should used here such a hybrid model.

We have to consider first elementary $\gamma p$ or $\gamma n$ 
cross sections.
The high energy component cross section (Tab.~\ref{tab:qd1}) on proton can be decomposed into a sum 
of three terms, named for brevity diffractive, Sullivan
and hadronization:
\begin{eqnarray}
\sigma_{\gamma p \to p}
&=& \sigma_{\gamma p \to p}^{diff}
 + \sigma_{\gamma p \to p}^{Sull.}
 + \sigma_{\gamma p \to p}^{hadr}      \; , \nonumber \\
\sigma_{\gamma p \to n}
&=& \sigma_{\gamma p \to n}^{diff}
 + \sigma_{\gamma p \to n}^{Sull.}
 + \sigma_{\gamma p \to n}^{hadr}      \; .  
\label{gammap_nucleon}
\end{eqnarray}
In an analogous way for production for scattering on neutron:
\begin{eqnarray}
\sigma_{\gamma n \to p}
&=& \sigma_{\gamma n \to p}^{diff}
 + \sigma_{\gamma n \to p}^{Sull.}
 + \sigma_{\gamma n \to p}^{hadr}      \; ,
\nonumber \\
\sigma_{\gamma n \to n}
&=& \sigma_{\gamma n \to n}^{diff}
 + \sigma_{\gamma n \to n}^{Sull.}
 + \sigma_{\gamma n \to n}^{hadr}      \; .  
\label{gamman_nucleon}
\end{eqnarray}
The decomposition in Eq.(\ref{gammap_nucleon}) and
Eq.(\ref{gamman_nucleon}) is written in the works
of one of the coauthors of the present paper 
\cite{SS1993,HSS1996,SBF1996,SBD1995}.
The diffractive components can be estimated as:
\begin{eqnarray}
 \sigma_{\gamma p \to p}^{diff} &\approx& 0.1 \; \sigma_{\gamma p \to p}
\; , \nonumber \\
 \sigma_{\gamma n \to n}^{diff} &\approx& 0.1 \; \sigma_{\gamma n \to n}
\; .
\end{eqnarray}
The value 0.1 is approximate contribution of diffractive processes $\gamma + p \to p + X$ i.e. processes with exchange of the pomeron. In this case the outgoing proton (or neutron) cannot receive a large kick of energy or momentum.

\begin{table}
\caption{Maximal contributions (in barns) of pre-equilibrium emission of $p$ and $n$
separately for different mechanisms.}
\begin{center}
\begin{tabular}{c|c|c|c|c|c}
\hline
&&&&&\\
$\sigma$~[b]    &     quasi-     &    nucl.    &    high & sum&ALICE \\
   &    deuteron     &    reson.    & energy   & & \cite{ALICE_protons, ALICE502a} \\&&&&&\\
\hline  &&&&&\\
1p   &   3.79    &  18.27   &  15.56& 37.62 & 40.$\pm$1.6\\
1n   &   3.79 -- 12.62    &  21.04   &  18.73& 43.56 --52.39&108.4$\pm$3.90\\
&&&&&\\
\hline
\end{tabular}\label{tab:qd1}
\end{center}
\end{table}

This means:
\begin{eqnarray}
\sigma_{AA \to p}^{diff} &\approx& 0.1\frac{Z}{A} \sigma_{AA}^{part} \; ,
\nonumber \\
\sigma_{AA \to n}^{diff} &\approx& 0.1\frac{N}{A} \sigma_{AA}^{part} \; .
\label{diffractive}
\end{eqnarray}
We consider diffractive processes on protons (with fraction $\frac{Z}{A}$ in the target nucleus) or on neutrons (with fraction $\frac{N}{A}$ in the target nucleus). Then the remaining diffractive components for $p \to n$ and $n \to p$
are small and can be ignored in our simple estimation.
For the so-called Sullivan processes
\footnote{The Sullivan processes are the only known to us 
explanation of the Gottfried Sum Rule violation
for deep inelastic scattering \cite{SS1993,HSS1996}.} 
one has
\begin{eqnarray}
\sigma_{\gamma p \to p}^{Sull.} &=& 
\sigma_{\gamma n \to n}^{Sull.} =
0.35 \; \frac{1}{3} \sigma_{\gamma p \to p}   \; , \nonumber \\
\sigma_{\gamma p \to n}^{Sull.} &=& 
\sigma_{\gamma n \to p}^{Sull.} =
0.35 \; \frac{2}{3} \sigma_{\gamma p \to p}   \; .
\end{eqnarray}
The 0.35 is an approximate probability to find (virtual) pion in the nucleon. Then reaction on quarks in the pion is called Sullivan process. The virtual pions in the nucleon allow to explain the Gotfried Sum Rule violation: $\int dx \frac{1}{x} \bigg [F_2^{ep}(x) -F_2^{en}(x) \bigg]\ne  \frac{1}{3} $ \cite{SS1993,HSS1996}. The $\frac{2}{3}$ is a relative probability to find charged pion and $\frac{1}{3}$ is a relative probability to register neutral pion. These two numbers reflect isospin structure of the couplings of pions to nucleons (see e.g. \cite{ericson}).

Combining the results for UPC
\begin{eqnarray}
\sigma_{AA \to p}^{Sull} &\approx&
\left(
\frac{Z}{A} 0.35 \frac{2}{3} + \frac{N}{A} 0.35 \frac{1}{3}
\right)
\sigma_{AA}^{part}
\; , \nonumber \\
\sigma_{AA \to n}^{Sull} &\approx&
\left(
\frac{Z}{A} 0.35 \frac{1}{3} + \frac{N}{A} 0.35 \frac{2}{3}
\right)
\sigma_{AA}^{part}
\; .
\label{Sullivan}
\end{eqnarray}

Assuming only light $u$, $d$ quarks and antiquarks
in $p$ and $n$ and $u(x) \propto d(x)$ (SU(2) symmetry
of quark distributions) we get 
\begin{eqnarray}
\sigma_{\gamma n}^{hadr} &=& \frac{2}{3} \sigma_{\gamma p}^{hadr} \;\; 
 \; ,  \nonumber \\
\sigma_{\gamma n}^{hadr} &=&  \sigma_{\gamma p}^{hadr} \;\;\; 
 \; 
\label{hadronization_inpandn}
\end{eqnarray}
for valence and sea dominance, respectively.
The contributions with $1p$ or $1n$ from that component is estimated as
\begin{eqnarray}
\sigma_{\gamma p \to p}^{hadr} &=& 0.7 \; \sigma_{\gamma p}^{hadr} \; ,
\nonumber \\
\sigma_{\gamma p \to n}^{hadr} &=& 0.3 \; \sigma_{\gamma p}^{hadr} \; ,
\nonumber \\
\sigma_{\gamma n \to n}^{hadr} &=& 0.7 \; \sigma_{\gamma n}^{hadr} \; ,
\nonumber \\
\sigma_{\gamma n \to p}^{hadr} &=& 0.3 \; \sigma_{\gamma n}^{hadr} \; .
\label{hadronizations_with_nucleon_inpandn}
\end{eqnarray}
These are typical numbers in hadronization models (see e.g.\cite{LEPTO}).
The hadronization component can be found by solving the set of
Eq.(\ref{hadronization_inpandn}) and
\begin{equation}
Z \cdot \sigma_{\gamma p}^{hadr} + N \cdot \sigma_{\gamma n}^{hadr}
\approx \sigma_{\gamma A}^{hadr}   \; ,
\label{hadronization_inA}
\end{equation}
where the hadronization component can be approximated as
\begin{equation}
\sigma_{\gamma A}^{hadr} \approx 0.55 \; \sigma_{\gamma A}^{part}  \; .
\end{equation}

The 0.55 is a typical probability of partonic effects in the hybrid models at high (virtual) photon energies.

No shadowing effects are included above. It would reduce 
somewhat our estimate.

Finally, we get for the hadronization component:
\begin{eqnarray}
\sigma_{AA \to p}^{hadr} &=&
\left(
\frac{Z}{A} 0.7 \cdot 0.55 + \frac{N}{A} 0.3 \cdot 0.55
\right)
\sigma_{AA}^{part}
\; , \nonumber \\
\sigma_{AA \to n}^{hadr} &=&
\left(
\frac{Z}{A} 0.3 \cdot 0.55 + \frac{N}{A} 0.7 \cdot 0.55
\right)
\sigma_{AA}^{part}
\; .
\label{hadronization_component}
\end{eqnarray}

The probability of baryon production in $\gamma p$ or $\gamma n$ collisions is 1, due to baryon number conservation. Mostly protons and neutrons are produced. In our approach we assume that $p \to p$ and $p \to n$ transitions 
happen in 100 \% and similarly for $n \to n$ and $n \to p$ transitions.
At high energies one may expect a small energy dependent reduction 
(less than 5 \%) due to hyperon production, which we neglect
in the current estimation.

Finally, combining the different components, we get
the following estimate:
\begin{eqnarray}
\sigma_{AA \to 1p} &=& 38 \; b   \; , \nonumber \\
\sigma_{AA \to 1n} &=& 44 \; b   \; .
\end{eqnarray}
In Table~\ref{tab:qd1}, we show a decomposition of the cross sections (in b)
into components from different regions/mechanisms. While the emission of proton from the quasi-deuteron was fitted to the data (see Fig.~\ref{fig:quasi_deuteron}), it is rather difficult to make a similar limitation for neutron emission from a quasi-deuteron. Instead, we show a minimal and maximal value of the cross section. The minimal value was assumed to be that found for proton emission. The cross section for neutron emission should be larger as there is no Coulomb barrier.

One can see that the biggest contributions come from the resonance
and partonic regions. Not all models on the market
include the partonic contributions. We note that it is impossible
to describe the ALICE data for proton production without
this high-energy component.

While for the one-proton emission the maximal cross section is compatible with the ALICE data \cite{ALICE_protons}, the situation for neutron emission is very different. It seems, the preequilibrium is a dominating mechanism of one-proton emission, in contrast the neutrons have very large equilibrium component (see \cite{Jucha2025}). Therefore the cross section for neutron emission above should not be compared with the ALICE neutron data \cite{ALICE502a}. It represents rather microscopic explanation of long energy tails in probability functions used in the TCM.

Please note that in the estimation in this section we have completely
neglected final state interactions (FSI) of produced and outgoing proton and neutron in 
the cold nuclear medium.
The rough agreement of our estimate with the ALICE data may mean that
potentially complicated nuclear effects in the intra-nuclear cascade
almost do not change the rate of $1p$ emissions.
The proton FSI is expected to reduce the population of one-proton channel.
On the other hand neutron FSI could enhance the population of one-proton 
channel.
The two effects work in opposite directions and tend to
compensate each other.
However, precise estimation of this effect is not easy.

The protons (and neutrons) formed in diffractive processes may
have too small energies to escape from nuclei,
therefore our estimate of $\sigma_{AA \to 1p}$ stays an upper limit 
for the single proton emissions.

The pre-equilibrium emission of one-proton with energy-dependent probability
\begin{equation}
P_p(\omega) \approx \sigma_{AA \to 1p}(\omega) / \sigma_{AA \to X}(\omega)
\label{probabilities} 
\end{equation}
starts a statistical cascade based on excited $^{207}$Tl$^{*}(E_{exc})$
while single neutron emission initiates the cascade based on excited 
$^{207}$Pb$^{*}(E_{exc})$.
The energy-dependent probabilities $P_p(\omega)$ and $P_n(\omega)$
could be extracted from Figs.~\ref{summary_emission}.
The chains mentioned above are initiated with probabilities:
$P_{^{207}{\textrm Tl}}(\omega) = P_{1p}(\omega)$ and
$P_{^{207}{\textrm Pb}}(\omega) = P_{1n}(\omega)$, which is about $\frac{1}{2}$.
We estimate that for $\omega >$ 200 MeV the excitation
energy of the initial $^{207}$Tl or $^{207}$Pb is typically
of the order of $E_{exc} \in $ (0, 30 MeV), which is relatively low.

However, it is rather difficult to calculate the initial distribution
of excitation energies $E_{exc}$ in $^{207}$Tl or $^{207}$Pb, 
respectively.

The current estimation of pre-equilibrium effects has
also interesting consequences for neutron emission.
If there were only pre-equilibrium emissions of neutrons
the $\frac{d \sigma}{d E_n}$ distribution in the fixed target case
($E_n$ is energy in the target rest frame) would be
approximately Gaussian \cite{Klusek-Gawenda:2013ema}.
The pre-equilibrium emission of neutrons causes appearance
of high-energy tails in the $\frac{d \sigma}{d E_n}$ distribution
observed in \cite{LEPRETRE1978} for fixed-target experiment
$\gamma$ + Pb.
Our estimate for one-neutron emission is that about 40 \%
of $1n$ emissions (40-45~b of the pre-equilibrium component compared
to 100-110 b measured by the ALICE Collaboration \cite{ALICE502a})
are in the tail. This seems to be roughly consistent
with the observation in \cite{LEPRETRE1981}.
Summarizing, the pre-equilibrium effects discussed here
are responsible for neutron high-energy tail observed long time ago
at Frascati and Saclay.

\subsection{Comment on $NN$ rescattering in 
the intra-nuclear cascade and short-range 
pn correlations}

In the last section, we discussed pre-equilibrium production of
protons and neutrons neglecting FSI effects.
The FSI may lead to elastic and inelastic collisions
in the nuclear matter. The nucleon hit by the photon is scattered and 
may interact with other nucleons in the target which will be rescattered.
There are the following combinations of such interactions
possible:\\
(1) $p(primary) + p(rescat.) \to 2p$,  \\
(2) $p(primary) + n(rescat.) \to p n$, \\ 
(3) $n(primary) + p(rescat.) \to p n$, \\
(4) $n(primary) + n(rescat.) \to 2n$.  \\
provided kinetic energies of nucleons are bigger than Fermi energy
due to Pauli exclusion principle. By primary we mean here proton
or neutron from $\gamma+N$ collision and by $rescat.$ proton or
neutron in the nucleus which take part in the rescattering process.
It was found in the last decade that the production of $pn$ pair is larger than $pp$ pairs \cite{proposal_GlueX,RCSCN2019,JLAB_rate_of_pncorrelations}.
Thus, at low energies, the production of $pn$ pairs is more probable than
the two others due to isospin structure of $NN$ interaction. This was not fully explained by theoretical works up to now.
Reaction (1) increases number of protons to two, so do not
participate to $1p$ channel but may be crucial for $2p$ channel.
Reaction (3) creates extra protons to $1p$ channel.

The $NN$ interaction is strongly energy dependent.
What are energies of hit nucleons?
Everything depends on the mechanism considered:\\
(A) Quasi-deuteron: low energy $p$ or $n$, \\
(B) Resonance region: intermediate energy $p$ or $n$,\\
(C) Partonic region: low (diffraction), intermediate (Sullivan)
    and high (nucleon fragmentation) energy of $p$ or $n$.\\
Reaction (A) may be also related to strong pn short range
correlations discussed recently (see e.g.\cite{proposal_GlueX}).
The region (B) was studied in HADES and COSY experiments.
The high energy region was not studied experimentally so far.

The consideration here shows that we can expect enhanced production
of $pn$ channel due to FSI and $pn$ short range correlations (SRC).

The ALICE collaboration measured also $pn$ channel 
(simultaneous use of proton and neutron Zero Degree Calorimeters (ZDCs) )
with the cross section 1.5~b \cite{ALICE_protons}, 
much lower than the cross section for inclusive single proton production.
This sets upper limit on FSI production of protons in 
neutron rescattering in $n+p \to n+p <$ 1.5 b.
The maximal cross section for $pn$ short range correlations (SRC) and/or FSI neutron effect
would be relatively low compared to our estimate
in the previous section. The estimated maximal effect of SRC 
one could think is 20\%\footnote{This is a rate of SRC correlated NN pairs \cite{RCSCN2019}
with assumed dominance of $pn$ correlations confirmed in 
\cite{JLAB_rate_of_pncorrelations}.} 
of 43~b of our Plane Wave Impulse Approximation (PWIA) prediction for single neutron production.
This estimate of SRC contribution ($\approx$ 8~b)
is much larger than the one measured by the ALICE collaboration.
Neutrons may be also produced in the equilibrium statistical emissions.
A measurement of ($1p 0n$) channel would be useful to shed more light
on the reaction mechanism. The data will be available now in Run3
with triggerless data taking mode.

\subsection{Emission of protons 
from the equilibrium phase}

Let us discuss now the emission of 1, 2 and 3 protons from
the equilibrium phase. It is rather difficult to
calculate the excitation energy of the residual nucleus.
In Fig.~\ref{fig:prot_probability_Estar} we show
probability of emission of 1, 2 and 3 protons from the excited
nucleus as a function of excitation energy, i.e.
energy after the pre-equilibrium emission of neutron or proton.
Here, for simplicity, we assume that the initial nucleus is $^{208}$Pb.

\begin{figure}
\includegraphics[width=9cm]{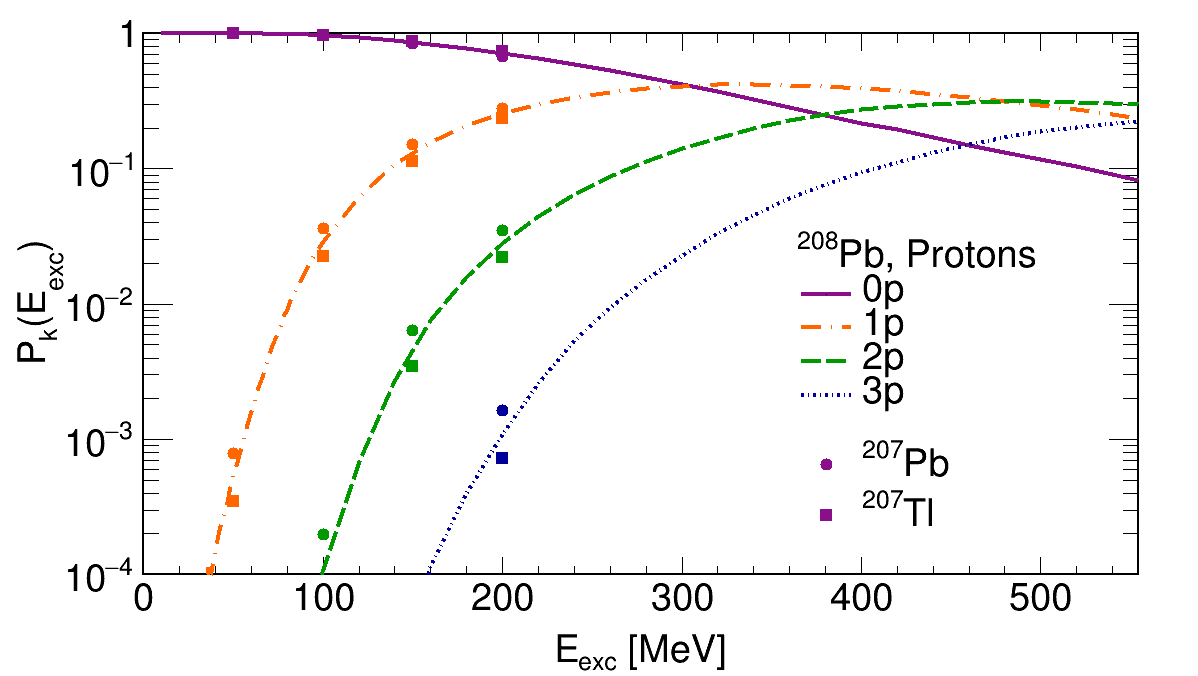}
\caption{Probability of emission of $1p$, 2p, 3p as
a function of excitation energy calculated within
GEMINI++ approach for $^{208}$Pb (lines). The circles are for initial $^{207}$Pb
and squares for initial $^{207}$Tl.
}
\label{fig:prot_probability_Estar}
\end{figure}

If the excitation energy of the nucleus is less than 100 MeV,
almost no protons are emitted. For energies greater than 200 MeV
, the emission of $1p$, $2p$ becomes sizable.
As discussed in previous section the emission of single proton is dominantly
due to pre-equilibrium emission.
However, it is very difficult to answer the question whether
the emission of $2p$ or $3p$ can be due to equilibrium emissions
or due to pre-equilibrium emission (FSI) discussed in
the previous section.

For $E_{exc}$ = 50, 100, 150, 200~MeV we show similar results
starting from $^{207}$Pb (circles) or $^{207}$Tl (squares).
There is only a small difference compared to the calculation
for $^{208}$Pb.
Our estimate suggests that typically 1 or 2 protons are emitted
in the pre-equilibrium phase (Fig.~\ref{mean_neutron}). 
When at higher photon energies nucleons are emitted in a preequilibrium process, this means that only a part of the initial photon energy is left in the nucleus. Then, residual nuclei are excited to rather small, typically $E_{exc} < $ 100~MeV. 

%
\section{Conclusion}

In the present paper, we have discussed the production of protons, neutrons, and nuclear isotopes in ultraperipheral collisions of
$^{208}$Pb + $^{208}$Pb.

Different combinations of pre-equilibrium and evaporation models
have been used. It has been very difficult to reproduce 
large cross section for one-photon production ($\sigma_{1p} \approx$ 40~b), 
as measured recently by the ALICE collaboration.

The only explanation is that the production of one proton happens
in a very broad range of photon energies in all possible 
photoproduction mechanisms such as emission of protons from:
interactions on quasi-deuterons, resonances excited 
on protons or neutrons or produced in scattering of 
photons on partons in individual nucleons.

The cross section for single proton emissions measured by the ALICE collaboration
is close to our estimate of the maximum possible cross section
from all above mechanisms, as discussed in a separate
section.
The GiBUU code, when combined with de-excitation models (GEMINI ++ or GEM2), leads to a large cross section for multipole protons production and a small
cross section for $1p$ detection which is not supported by the ALICE data.

We have also discussed the difference in cross sections for production
of $1p$, $2p$, $3p$, etc and that for production of Pb, Tl, Hg, Au, ...
isotopes. The equality of the corresponding cross sections {\bf (
    $\sigma_{1p} = \sigma_{Tl}$, $\sigma_{2p} = \sigma_{Hg}$,
    $\sigma_{3p} = \sigma_{Au}$, etc.)}
is assumed by the ALICE collaboration. 
However, it is different in models with sizable emission of
other charged particles such as $d$, $t$, $^{3}$He or $\alpha$ particles
such as HIPSE or EMPIRE. For example, a strong $\alpha$ particle emission would invalidate these equalities.
It is rather difficult to test these models with present
experimental infrastructure at the LHC where only neutrons or protons
can be measured.

No existing models, including TCM, EMPIRE, GIBUU, can explain
huge cross section for $1p$ emission as measured with the help of 
the proton zero-degree calorimeter.
We have presented a careful estimation of maximal possible cross section
for $1p$ emission and found that the corresponding ALICE data is very
close to this estimate. 
The new ALICE data for proton production in UPC strongly suggests that
single proton ejection is caused dominantly by pre-equilibrium processes
initiated by photons interacting with individual nucleons in
``colliding'' nuclei.

The emission of protons in $\gamma+^{208}$Pb collisions was not well
explored so far, especially at higher energies.
One possible place to perform such studies, at relatively high photon
energies, is JLab. Another option in more distant future is EIC
for $e+^{208}$Pb collisions (virtual photons).
It would be be good to plan such experiment(s) already now in order
to prepare underlying infrastructure.

\vspace{1cm}

{\bf Acknowledgments}

We are indebted to Chiara Oppedisano for exchange of information on details of the ALICE ZDC measurements, Or Hen for a discussion about possibility
to measure $\gamma + ^{208}$Pb$ \to p$ process at JLab,
Jan Ryckebusch for exchange of general information about $\gamma + A$
processes, Irene Dedes for preparation
Fig.\ref{fig:single_particle_potentials} 
for displaying neutron and proton single particle potentials for 
$^{208}$Pb, Mariola K\l{}usek-Gawenda for discussing several aspects
of UPC and Micha\l{} Ciema\l{}a for finding us a unique reference for the
$^{208}$Pb$(\gamma,p)$ reaction and technical help with GEMINI++ calculations.



\end{document}